\documentclass[journal]{IEEEtran}
\IEEEoverridecommandlockouts
\usepackage[tbtags]{amsmath}
\usepackage{comment}
\usepackage{color}
\usepackage{graphicx}
\usepackage{subfigure}
\usepackage{CJK}
\usepackage{extarrows}
\usepackage{threeparttable}
\usepackage{graphicx}
\usepackage{epstopdf}
\usepackage{amsthm}

\newtheorem{prop}{Proposition}
\newtheorem{thre}{Theorem}
\newtheorem{remark}{Remark}
\newtheorem{coro}{Corollary}
\usepackage{bm}
\usepackage{amssymb}
\usepackage{float}
\usepackage{multirow}
\usepackage{stfloats}
\usepackage{cite}
\usepackage{algorithm}
\usepackage{algpseudocode}
\usepackage{verbatim}
\usepackage{tikz}
\usepackage{array}

\def\BibTeX{{\rm B\kern-.05em{\sc i\kern-.025em b}\kern-.08em
    T\kern-.1667em\lower.7ex\hbox{E}\kern-.125emX}}
\begin{document}
\title{Orthogonal Delay-Doppler Division Multiplexing Modulation with Tomlinson-Harashima Precoding
\author{Yiyan~Ma,~\IEEEmembership{Member,~IEEE,}
        Akram~Shafie,~\IEEEmembership{Member,~IEEE,}
        Jinhong~Yuan,~\IEEEmembership{Fellow,~IEEE,}\\
        Guoyu~Ma*,~\IEEEmembership{Member,~IEEE,}
        Zhangdui~Zhong,~\IEEEmembership{Fellow,~IEEE,}
        ~and~Bo~Ai*,~\IEEEmembership{Fellow,~IEEE}
\thanks{Yiyan Ma, Guoyu Ma, Zhangdui Zhong and  Bo Ai are with the School of Electronic and Information Engineering, Beijing Jiaotong University, Beijing, 100044, China. Akram Shafie and Jinhong Yuan are with the School of Electrical Engineering and Telecommunications, University of New South Wales, Sydney, NSW 2052, Australia. A preliminary version of this work was accepted by 2024 IEEE Global Communications Conference (Globecom). (Corresponding authors: boai@bjtu.edu.cn, magy@bjtu.edu.cn).}
\thanks{This paper is supported by the National Natural Science Foundation of China under Grant 62221001, 62341127, 62471025 and 62101024; the Beijing Natural Science Foundation (L234083); the National Key Research and Development Program (2021YFB2900301, 2020YFB1806604, 2021YFB3901302); and the Fundamental Research Funds for the Central Universities (2022JBQY004, 2022JBXT001, 2024JBMC004).}
}}
\maketitle

\begin{abstract}
The orthogonal delay-Doppler (DD) division multiplexing (ODDM) modulation has been recently proposed as a promising modulation scheme for next-generation communication systems with high mobility. Despite its benefits, ODDM modulation and other DD domain modulation schemes face the challenge of excessive equalization complexity. To address this challenge, we propose time domain Tomlinson-Harashima precoding (THP) for the ODDM {transmitter}, to make the DD domain single-tap equalizer feasible, thereby reducing the equalization complexity. In our design, we first pre-cancel the inter-symbol-interference (ISI) using the linear time-varying (LTV) channel information. Second, different from classical THP designs, we introduce a modified modulo operation with an adaptive modulus, by which the joint DD domain data multiplexing and time-domain ISI pre-cancellation can be realized without excessively increasing the bit errors. We then analytically study the losses encountered in {this} design, namely the power loss, the modulo noise loss, and the modulo signal loss. Based on {this} analysis, BER lower bounds of the ODDM system with time domain THP are derived when 4-QAM or 16-QAM modulations are adopted for symbol mapping in the DD domain. Finally, through numerical results, we validate our analysis and then demonstrate that the ODDM system with time domain THP is a promising solution to realize better BER performance over LTV channels compared to orthogonal frequency division multiplexing systems with single-tap equalizer and ODDM systems with maximum ratio combining.
\end{abstract}

\begin{IEEEkeywords}
Orthogonal delay-Doppler division multiplexing, Tomlinson-Harashima precoding, single-tap equalizer.
\end{IEEEkeywords}

\section{Introduction}
As the highly mobile communication scenarios emerge, the orthogonal frequency division multiplexing (OFDM) modulation scheme in the fifth-generation (5G) communications system is challenged by the inter-carrier-interference (ICI) caused by the Doppler shift and the fast and deep fading of the channel transfer function (CTF) \cite{zhiqiang,weijiesurvey}. {In recent years}, delay-Doppler (DD) domain multicarrier (DDMC) modulation schemes are proposed to address {this} challenge \cite{zhiqiang}. Compared to OFDM scheme, the DDMC schemes multiplex data in the DD domain rather than the time-frequency (TF) domain. Due to this, and as DDMC schemes rely on the stationary period and bandwidth of the channel spreading function (CSF) rather than the fast-fading CTF, they are able to explore the full channel diversities, thus enabling reliable communications in {\color{black}high-speed mobile} scenarios \cite{bb2}.\par


The well-known orthogonal time frequency space (OTFS) modulation scheme was among the first introduced DDMC schemes \cite{bb2}. With DD domain data multiplexing and data detection, it realizes better transmission reliability compared to the OFDM scheme over the same linear time-varying (LTV) channels \cite{bb4,ziji2}. Despite its benefits, there are several unresolved challenges associated with OTFS, particularly concerning its out-of-band emission (OOBE) \cite{b3,Akram}. {In details}, on one hand, the practically realizable pulse-shaping and filters were not well considered. Instead, the rectangle pulse with severe OOBE was usually adopted \cite{CS}. On the other hand, the OTFS modulation based on other OOBE-limited practical pulses, such as the prolate spheroidal pulses, was proposed; but it demonstrated bit error rate (BER) performances greatly worse than that with the impractical rectangular pulses \cite{bb5}. To tackle these challenges, the orthogonal DD division multiplexing (ODDM) modulation was proposed in \cite{b3,b3b}. Therein, the transceiver pulses which were orthogonal with respect to (w.r.t.) the DD domain resolutions were proposed, namely the DD plane orthogonal pulse (DDOP). Since the DDOP was designed based on the square-root Nyquist (SRN) sub-pulses which acted as low-pass filters at transceivers, OOBE was successfully suppressed in the ODDM scheme.\par



Although the ODDM scheme is promising for practical implementation compared to the OTFS scheme, it is still faced with the challenge of the excessive {equalization or receiver} complexity \cite{b3,b3b}. In all the DDMC schemes, i.e., ODDM as well as OTFS schemes, the received symbols {suffer from} inter-symbol-interference (ISI) over the LTV channel \cite{b3}. In particular, due to multipath effects, each received symbol in DD domain is composed of multiple transmitted DD domain symbols. Due to this, current ODDM systems utilize equalizers with high complexity, unlike the low-complexity single-tap equalizer employed by OFDM systems over linear time-invariant (LTI) channels. These include the minimum mean square error (MMSE) equalizer \cite{Kehan}, message passing (MP) algorithm-based equalizer \cite{bb4}, and the maximum ratio combining (MRC) equalizer \cite{bb15,luyang}.\par
 Motivated by the need to perform low complex equalization, we propose to introduce {and leverage} Tomlinson-Harashima precoding (THP) to the ODDM systems. In THP, the ISI is pre-cancelled based on the known channel information at the transmitter, thereby making a single-tap equalizer feasible at the receiver \cite{b6,b7}. Additionally, THP employs a modulo operation to constrain the power of the transmitted symbols \cite{b6,b7}. Due to the benefits of THP of low implementation complexity, compatibility with transmitter power amplifiers, and not causing error propagation \cite{b4}, THP has been widely investigated in communication systems over the LTI channels, e.g., the digital subscriber line systems \cite{b8}, multi-user multiple-input-multiple-output systems \cite{A3}, and the fiber-optic systems \cite{blight}.\par
 
  Despite the efforts in \cite{b8,A3,blight}, the applicability of THP for TF domain multiplexing systems over LTV channels is very limited since the CTF is fast-fading, potentially leading to suboptimal performance for TF domain multiplexing systems with THP. Nevertheless, for DDMC systems designed based on the stationary CSF rather than the fast-fading CTF, the transmitter can access the CSF as it remains stationary with high-mobility. Consequently, it becomes feasible to incorporate THP into ODDM systems over LTV channels.\par


{Note that} designs on DDMC systems with THP are very limited. Very recent works \cite{b9,ba} designed OTFS systems with THP. In \cite{b9} and \cite{ba}, the THP was performed in the DD domain, where certain DD domain symbols were kept as null to implement ISI pre-cancellation. Despite the potential of the designs in \cite{b9} and \cite{ba}, setting null symbols could cause complex loops and lead to significant resource overhead.\par

In this work, we propose time domain THP for the ODDM systems to enable reliable and low overhead communications in highly mobile scenario. The main contributions are summarized as follows:
\begin{itemize}
  \item We propose {an ODDM system} with time domain THP. Owing that THP is performed in the time domain rather than the DD domain, the proposed design avoids the high resource overhead and precoding complexity as in previous works \cite{b9,ba}. 
  \item Since our ODDM system with time domain THP design necessitates DD domain symbol mapping and time domain THP over LTV channels, the THP operations adopted in classical THP designs \cite{b6,b7} cannot be directly used in our design. Thus, we propose ISI pre-cancellation in the time domain that leverages the features of DD domain LTV channel within the stationary period. Second, we propose a modified modulo operation with an adaptive modulus to mitigate the excess system performance loss.
  \item We provide extensive analysis of the bit error rate (BER) performance for our design when 4-QAM and 16-QAM modulations are adopted for {DD domain symbol mapping}. In particular, we decompose the performance loss that our design encounters into three, namely, {\it power loss}, {\it modulo noise loss}, and {\it modulo signal loss}. First, the reasons for these losses to occur in our proposed design are studied. Second, BER associated with these losses are derived analytically. Finally, BER lower bounds for our ODDM system with time domain THP are obtained for both 4-QAM and 16-QAM modulations.
      
  \item Through numerical simulations, we first validate our BER analysis. Second, we demonstrate the superiority of the proposed designs compared to the OFDM systems with single-tap equalizer and the ODDM system with MRC equalizer. In particular, we show that the proposed design (i) demonstrates superior performance in the high signal to noise ratio (SNR) region or when the channel diversity is limited, and (ii) does not introduce an error floor.
\end{itemize}

The rest of this paper is structured as follows: The standalone ODDM system is briefly introduced in Section \uppercase\expandafter{\romannumeral2}. The proposed ODDM system with time domain THP is detailed in Section \uppercase\expandafter{\romannumeral3}. The analytical results of the BER performance losses of the proposed design are discussed in Section \uppercase\expandafter{\romannumeral4}. The numerical results are presented in Section \uppercase\expandafter{\romannumeral5}, and, finally, the conclusion and future works are given in Section \uppercase\expandafter{\romannumeral6}.\par
\section{The Standalone ODDM System}
\begin{figure*}
  \centering
  \subfigure[]{
  \includegraphics[height = 60pt,width=.6\textwidth]{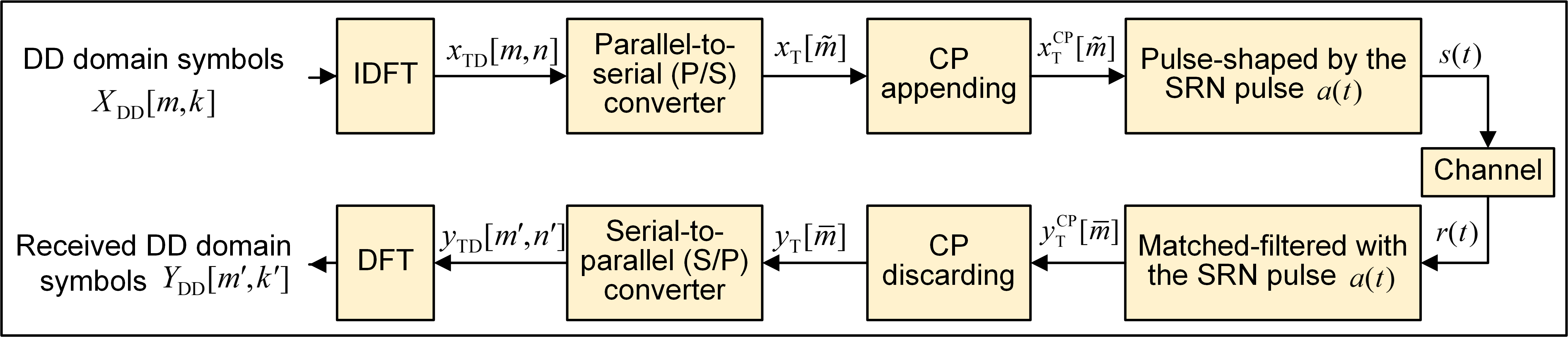}
  \label{FIG2B}}
  \subfigure[]{
  \includegraphics[height = 60pt, width=.35\textwidth]{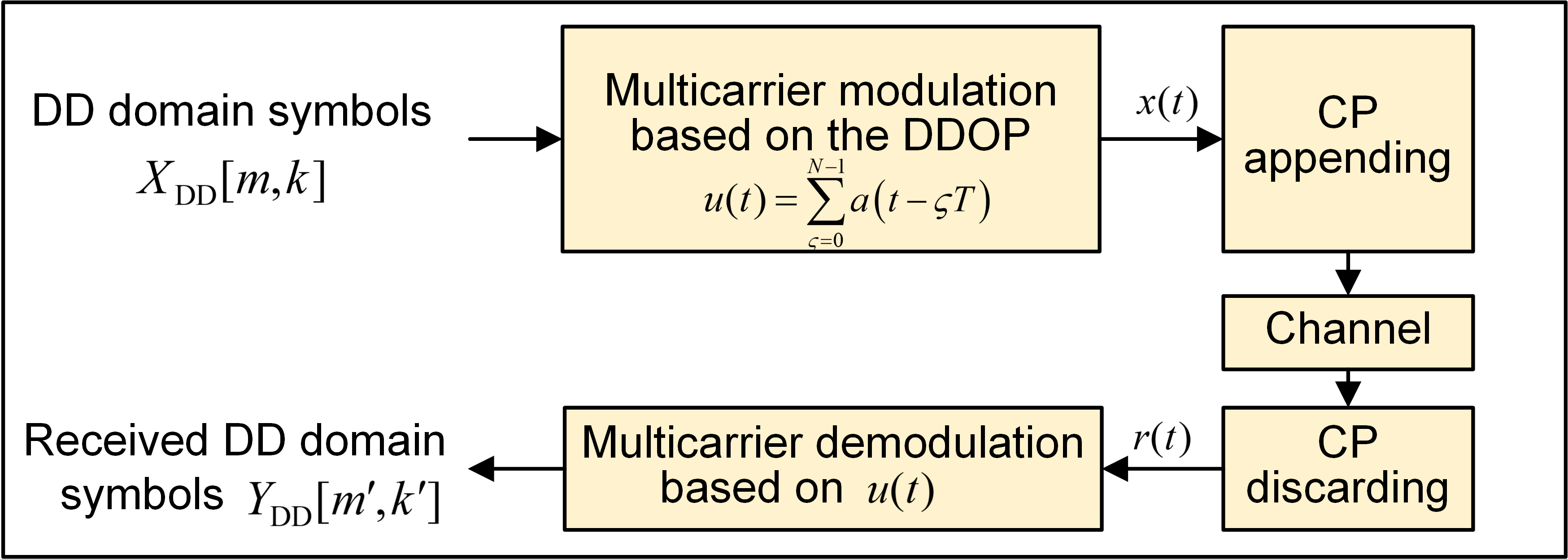}
  \label{FIG2A}}
    \vspace{-0.5em}
  \caption{Illustration of the ODDM system based on (a) the sample-wise pulse-shaping with $a(t)$ and (b) multicarrier modulation with $u(t)$.}
  \vspace{-1em}
  \end{figure*}

{
In this section, {we briefly explain an ODDM system \cite{b3,b3b} for easy presentation of later sections}. A block diagram of the standalone ODDM system is shown in Fig. \ref{FIG2B}.
\subsection{Transmitter}
We assume the ODDM system occupies the bandwidth $M\Delta f$ and the time duration $NT$, where $M$ and $N$ are the number of symbols in the delay and Doppler domains in an ODDM frame, respectively, {and $T =\frac{1}{\Delta f}$.} Denote the DD domain $\cal M$-QAM modulated data as ${X}_{\rm DD}[m,k]$ with $m \in \{0,1,...,M-1\}$ and $k \in \{0,1,...,N-1\}$. By conducting the inverse discrete Fourier transform (IDFT) on $X_{\rm DD} [m,k]$, the time-delay domain symbols $x_{\rm TD}[m,n]$ can be obtained as:
\begin{equation}
x_{\rm TD}[m,n] = \frac{1}{{\sqrt N }}\sum\limits_{k = 0}^{N - 1} X_{\rm DD} [m,k]{e^{j2\pi \frac{{kn}}{N}}},
\label{Equ2}
\end{equation}
where $n \in \{0,1,...,N-1 \}$. Subsequently, $x_{\rm TD}[m,n]$ is sent through the parallel-to-serial (P/S) converter to obtain the time domain digital sequence $x_{\rm T}[\tilde m]$, where $\tilde m \in \{0,1,...,NM-1\}$ and $x_{\rm T}[nM+m] = x_{\rm TD}[m,n]$. Next, a cyclic prefix (CP) with $L_{\rm CP}$ symbols is appended to $x_{\rm T}[\tilde m]$ to get $x^{\rm CP}_{\rm T}[\tilde m]$, {\color{black} where $\tilde m \in \{-{L_{\rm CP}},-{L_{\rm CP}}+1,...,NM-1\}$. Therein, $x^{\rm CP}_{\rm T}[\tilde m] = x^{\rm CP}_{\rm T}[\tilde m+M]$ holds for the CP symbols with $-L_{\rm CP}\le\tilde m <0$. Besides, $L_{\rm CP} \ge \tau_{P}\frac{M}{T}$ is supposed where} $\tau_P$ is the maximum {delay spread} of propagation paths. Finally, $x^{\rm CP}_{\rm T}[\tilde m]$ is convoluted with the SRN pulse $a(t)$ to obtain the time domain analog signal $s(t)$ for transmission:
\begin{equation}
s(t) = \sum\limits_{\tilde m = -L_{\rm CP}}^{NM - 1} {{x^{\rm CP}_{\rm T}[\tilde m]} } a\left( {t - \tilde m\frac{T}{M}} \right).
\label{Equ3}
\end{equation}
Therein, the span of $a\left( {t} \right)$ is $2{\cal Q}$ symbol durations, i.e., $a\left( {t} \right){\ne} 0$ for $t {\in}[-{\cal Q} \frac{M}{T}, {\cal Q} \frac{M}{T}]$ and $a\left( {t} \right) =  0$ otherwise, where ${\cal Q}$ is an integer.\par
\begin{remark}
There exist two approaches to implement ODDM, namely the multicarrier modulation-based approach shown in Fig. \ref{FIG2A} and the symbol-wise pulse-shaping-based approach discussed above and shown in Fig. \ref{FIG2B} \cite{b3,b3b}. It has been proven that in the multicarrier modulation-based approach, the ODDM signal can attain strict bi-orthogonality w.r.t. the DD domain resolutions \cite{b3}. Meanwhile, the symbol-wise pulse-shaping-based approach has been demonstrated to provide only near-perfect approximation bi-orthogonality w.r.t the DD domain resolutions. But, it provides a simpler implementation approach. Considering this benefit of simple implementation, in this work, the symbol-wise pulse-shaping-based ODDM implementation is adopted.
\end{remark}
\subsection{Linear Time-varying (LTV) Channel}
In high mobility scenarios, the CSF can be modeled as{ \cite{matz}}:
\begin{equation}
h(\tau,\nu) = \sum^P_{p=1} h_p \delta(\tau-\tau_p)\delta(\nu-\nu_p),
\label{Equ4}
\end{equation}
where $\tau$ and $\nu$ denote the delay and Doppler {variables}, respectively, $P$ is the total number of propagation paths, and $h_p$, $\tau_p$, and $\nu_p$ denote the channel fading, delay shift, and Doppler shift of the $p$th propagation path, respectively. Following \cite{bb4}, $h_p$, $\tau_p$, $\nu_p$, and $P$ are considered to be invariable during $2NT$ based on the stationarity of the CSF. For simplicity and following \cite{b3,b3b}, we assume that delay and Doppler shifts of the propagation paths takes on-grid values. This leads to $\tau_p = l_p \frac{T}{M}$ and $\nu_p = k_p \frac{1}{NT}$ with integers $l_p$ and $k_p$.\footnote{For practical systems, the delays and Doppler shifts of the propagation paths may not be on-grid. However, the delays and Dopplers of the equivalent channel obtained after filtering and sampling would be on-grid \cite{JUN}.} Without loss of generality, we consider $\tau_1 \le \tau_2 \le...\le \tau_P$ and $\tau_1  = 0$.
\subsection{Receiver}
When the signal is propagated through the LTV channel, the received signal $r(t)$ is obtained as:
\begin{equation}
r(t) = \sum^{P}_{p=1} h_{p} s(t-\tau_p) e^{j2\pi\nu_p(t-\tau_p)} + w(t),
\label{Equ5}
\end{equation}
where $w(t)$ is the additive white Gaussian noise (AWGN). At the receiver, the $r(t)$ is first send through an $a(t)$-based matched filter to obtain:
\begin{equation}
{y}(t) = \int {{r}(t')} a\left( {t' - t} \right)dt'.
\label{Equ6}
\end{equation}
Then ${y}(t)$ is sampled at $t = \bar m\frac{T}{M}$ to obtained the discrete time domain signal $y^{\rm CP}_{\rm T}[\bar m]$, with $\bar m\in \{-L_{\rm CP},...,0,1,...,NM-1\}$. Based on (\ref{Equ3}), (\ref{Equ5}), and (\ref{Equ6}), $y^{\rm CP}_{\rm T}[\bar m]$ can be expanded as:
\begin{equation}
\begin{aligned}
&y^{\rm CP}_{\rm T}[\bar m] = {y}(t)|_{t=\bar m\frac{T}{M}}= \sum^{P}_{p = 1}{h_p}{e^{- j2\pi {\nu _p}{\tau _p}}} {\sum\limits_{\bar m = -L_{\rm CP}}^{NM - 1} {x^{\rm CP}_{\rm T}[\tilde m]} }\times \\
&{\int\limits_{{\tau _p}}^{NT+\tau _p} {{e^{j2\pi {\nu _p}t'}} a\left( {{t' {-} {\tau _p}} {-} \tilde m\frac{T}{M}} \right)}a\left( {t' {-}  \bar m\frac{T}{M}} \right)dt'} + w[\bar m].
\end{aligned}
\label{Equ8}
\end{equation}
When the support of the SRN pulse $a(t)$, $[-{\cal Q}\frac{T}{M},{\cal Q}\frac{T}{M}]$, is limited, i.e., ${\cal Q}\ll M$, the phase shift term ${e^{j2\pi {\nu _p}t'}}$ in the integral in (\ref{Equ8}) can be regarded as approximately invariable over time. Considering this, the integral in (\ref{Equ8}) can be approximated to 1 for $\bar m = \tilde m + \tau_p$ and $t' = \bar m\frac{T}{M}$, and $0$ otherwise \cite{b3}. Based on this and disregarding the CPs in $y^{\rm CP}_{\rm T}[\bar m]$ (i.e., the first $L_{\rm CP}$ symbols of $y^{\rm CP}_{\rm T}[\bar m]$), the discrete time domain signal $y_{\rm T}[\bar m]$ can be obtained as:
\begin{equation}
\begin{aligned}
 & y_{\rm T}[\bar m]= \sum^{P}_{p = 1}{h_p}{e^{j2\pi {\nu _p}(\bar m\frac{T}{M} -\tau_p)}}{x_{\rm T}\left[[\bar m -l_p]_{NM}\right]} + w[\bar m].
\end{aligned}
\label{Equ9BXXX}
\end{equation}
Therein, $\bar m\in \{0,1,...,NM-1\}$, $[x]_{NM}$ refers to the modulo operation with modulus $NM$, and is given by $[x]_{NM} = x - NM\lfloor\frac{x}{NM}\rfloor$ with $\lfloor\cdot\rfloor$ denoting the floor operation. Then $y_{\rm T}[\bar m]$ is send through the serial-to-parallel (S/P) converter to obtain:
\begin{equation}
\begin{aligned}
y_{\rm TD}[m' ,n']=&\sum^{P}_{p = 1}{h_p}{e^{j2\pi {\nu _p}(m'\frac{T}{M} + n'T-\tau_p)}}{x_{\rm TD}[\hat m^{\prime}_p,\hat n^{\prime}_p]}\\ &+ w[m',n'],
\end{aligned}
\label{Equ9}
\end{equation}
where 
\begin{equation}
\begin{aligned}
& {x_{\rm TD}[\hat m^{\prime}_p,\hat n^{\prime}_p]} = \begin{cases}{\begin{aligned}& {x_{\rm TD}[m'-l_p,n']},\end{aligned}} & m'\ge l_p,\\
{\begin{aligned}& {x_{\rm TD}[[m'-l_p]_M,[n'-1]_N]},\end{aligned}} &  m'<l_p,
\end{cases}
\end{aligned}
\end{equation}
with $y_{\rm TD}[m',n']=y_{\rm T}[n'M+m']=y_{\rm TD}[\bar m]$, $n' \in \{0,1,...,N-1\}$, and $m' \in \{0,1,...,M-1\}$. Finally, $y_{\rm TD}[m',n']$ is transformed into DD domain symbols $Y_{\rm DD}[m',k']$ through the DFT operation:
\begin{subequations}
\begin{equation}
\begin{aligned}
& Y_{\rm DD}[m',k']  = \frac{1}{\sqrt{N}}\sum^{N-1}_{n'=0}y_{\rm TD}[m',n']e^{-j2\pi\frac{n'k'}{N}},\\
& {=} {\begin{cases}\begin{aligned} &\frac{1}{{\sqrt N }}\sum\limits_{n' = 0}^{N - 1} {{h_p}} {e^{ - j2\pi \frac{{n'k'}}{N}}}{e^{j2\pi {\nu _p}(m'\frac{T}{M} + n'T - {\tau _p})}}\\ & \times {x_{{\rm{TD}}}}[m' - l_p,n']+ w[m',k'],\end{aligned}&{m' \ge {l_p}},\\
{\begin{aligned} &\frac{1}{{\sqrt N }}\sum\limits_{n' = 0}^{N - 1} {{h_p}} {e^{ - j2\pi \frac{{n'k'}}{N}}}{e^{j2\pi {\nu _p}(m'\frac{T}{M} + n'T - {\tau _p})}}\\ &\times {x_{{\rm{TD}}}}[m' - l_p + M,[n' - 1]_N]+ w[m',k'],\end{aligned}}&{m' < {l_p}}.
\end{cases}}\\
\end{aligned}
\label{Equ12a}
\end{equation}
\begin{equation}
\begin{aligned}
& {=} \sum^{P}_{p = 1}{h_p}{e^{j2\pi {\nu _p}(m'\frac{T}{M} - {\tau _p})}}{{{\tilde X}_{{\rm{DD}}}}[{\hat m'_p},{\hat k'_p}]}+ w[m',k'],
\end{aligned}
\label{Equ12}
\end{equation}
\end{subequations}
where (\ref{Equ12a}) and (\ref{Equ12}) are obtained by substituting (\ref{Equ2}) in (\ref{Equ12a}) and (\ref{Equ9}) in (\ref{Equ12}), respectively. In (\ref{Equ12}), ${{{\tilde X}_{{\rm{DD}}}}[{\hat m'_p},{\hat k'_p}]}$ is:
\begin{equation}
\begin{split}
& \!\!\!{{{\tilde X}_{{\rm{DD}}}}[{\hat m'_p},{\hat k'_p}]} = \\
& \!\!\!\!\!\begin{cases} {\begin{aligned}&{{X_{{\rm{DD}}}}[[m' - l_p]_M,{{\left[ {k' - k_p} \right]}_N}]},\end{aligned}} \,\,\,\,\,& m'\!\ge\! l_p,\\
{{e^{ \!-\! j2\pi \frac{{{{\!\left[ {k' \!-\! k_p}\right]}_N}}}{N}}}{{X_{{\rm{DD}}}}[[m' - l_p]_M,{{\left[ {k' - k_p} \right]}_N}]},} \,\,\,\,\,& m'\!<\!l_p,
\end{cases}
\end{split}
\end{equation}
with $k' \in \{0,1,...,N-1\}$. Additionally, we note that $w[m',k']$ in (\ref{Equ12}) is also an AWGN variable with the same distribution as $w[m',n']$ in (\ref{Equ9}), due to the fact that the complex Gaussian variables are isotropic after the DFT operation \cite{b2}.\par

When $Y_{\rm DD}[m',k']$, $X_{\rm DD}[m,k]$, and $w[m',k']$ in (\ref{Equ12}) are vectorized, the DD domain input-output relation can be obtained in the matrix-vector form as:
\begin{equation}
\begin{aligned}
{{\bf{Y}}_{{\rm{DD}}}} = {{\bf{H}}_{{\rm{DD}}}}{{\bf{X}}_{{\rm{DD}}}} + {\bf w},
\end{aligned}
\label{Equ13}
\end{equation}
where ${\bf X}_{\rm DD}, {\bf Y}_{\rm DD} \in {\mathbb C}^{NM\times 1}$, and ${\bf w} \in {\mathbb C}^{NM\times 1}$ are the vectorized DD domain input, output, and AWGN noise, respectively, with ${\bf X}_{\rm DD}(m+kM+1) = X_{\rm DD}[m,k]$, ${\bf Y}_{\rm DD}(m'+k'M+1) = Y_{\rm DD}[m',k']$, and ${\bf w}(m'+k'M+1)\sim {\cal {CN}}(0,\sigma^2_w)$. Moreover, ${{\bf{H}}_{{\rm{DD}}}} \in {\mathbb C}^{NM\times NM}$ is the DD domain equivalent channel matrix, whose elements are given by:
\begin{equation}
\label{DDCHANNEL}
 \begin{aligned}
 & {{\bf{H}}_{{\rm{DD}}}}(m'+k'M+1,[m' - {l_p}]_M+{{\left[ {k' - {k_p}} \right]}_N}M+1)\\
 &  = \begin{cases}
 {h_p}{e^{j2\pi {\nu _p}(m'\frac{T}{M} - {\tau _p})}}, & m'\ge l_p,\\
 {h_p}{e^{j2\pi {\nu _p}(m'\frac{T}{M} - {\tau _p})}}{e^{ - j2\pi \frac{{{{\left[ {k' - {k_p}} \right]}_N}}}{N}}}, & {\rm else}.
 \end{cases}
 \end{aligned}
\end{equation}
\subsection{Equalization}

At the receiver, {${{\bf{X}}_{{\rm{DD}}}}$ needs to be estimated based on ${{\bf{Y}}_{{\rm{DD}}}}$ and ${{\bf{H}}_{{\rm{DD}}}}$ via an equalizer}. As can be observed in (\ref{DDCHANNEL}), ${{\bf{H}}_{{\rm{DD}}}}$ is not a diagonal matrix, which shows that standalone {ODDM system} experiences ISI. Moreover, ${{\bf{H}}_{{\rm{DD}}}}$ in (\ref{DDCHANNEL}) cannot be simply diagonalized through the IDFT and DFT operations, as performed in OFDM systems \cite{Kehan}. Due to these, the equalization complexity of standalone ODDM systems can be exceedingly high \cite{Kehan}. In particular, traditional linear equalizers, which rely on principles such as MMSE to compute the inverse of the DD domain equivalent channel matrix ${\bf H}_{\rm DD}$, exhibit a computational complexity of ${\cal O}(N^3M^3)$ \cite{Chock}. This complexity becomes particularly burdensome when $MN$ is large. Meanwhile, {iteration-based equalizers}, such as the MP detection strategy outlined in \cite{bb4} and the MRC detection proposed in \cite{bb15}, which still have high complexity, but not that high as MMSE, have been proposed for ODDM systems. However, it is worth noting that the performance of these {iteration-based equalizers} may be constrained by error propagations \cite{Kehan}.\par

{\color{black}To overcome these challenges, in this work, we explore ODDM systems with time domain THP based on the assumption that the transmitter is aware of the perfect CSF. As a result, the single-tap equalizer could be made possible for the ODDM systems.\par}
{\color{black}
  \begin{remark}
  {\color{black}We note that in high-speed mobile scenarios, it is still possible for the transmitter to obtain the CSF $h(\tau,\nu)$. For example, the CSF for a given ODDM frame (denoted as the $i$-th frame) can be approximated in three steps \cite{weijie,shuangyangisac}. For ease of illustration, let us consider the channels where the signal undergoes transmitter-receiver and transmitter-receiver-transmitter as the one-way and two-way channels, respectively. As the first step, using the pilot information, the two-way channel will be determined at the transmitter for the $(i-1)$-th ODDM frame, where the ODDM frame duration $NT$ will be carefully selected such that DD domain representation of the channel remains static for the duration $2NT$ \cite{ziji1}. Second, based on the estimated relationship between the two-way and one-way channels, one-way channel parameters for the $(i-1)$-th frame could be estimated. Third, using the one-way channel parameters of the $(i-1)$-th frame, and the rate at which channel parameters are changing over frames, one-way channel parameters for the $i$-th frame will be approximated. Because of the availability of such approaches to estimate the CSF at the transmitter, and following many prior studies for DDMC where the availability of CSF at the transmitter is considered \cite{bb4,bb15,b9,Shuangyang1,Shuangyang2,SKM}, in this work, we focus on designing our ODDM scheme with time domain THP based on the perfect CSF assumption at the transmitter.}
  
  \end{remark}}
\section{ODDM System with Time Domain THP}
In this section, the ODDM system with time domain THP is introduced. We note that THP pre-cancels ISI successively at the transmitter \cite{b6,b7} based on the lower triangle structure of the channel matrix {in the time domain}, similar to the decision-feedback equalizer (DFE) at the receiver \cite{bb15}. However, for the standalone ODDM system, its equivalent DD domain channel matrix ${\bf H}_{\rm DD}$ given in (\ref{DDCHANNEL}) does not adhere to the lower triangular form. Due to this and the fact that the equivalent time domain channel matrix of the standalone ODDM system adhere to the lower triangular form, we introduce {\it THP in the time domain} rather than the DD domain. The proposed ODDM system with time domain THP is shown in Fig. \ref{FIG3}. \par


\begin{figure*}[t]
  \centering
  \includegraphics[width=\textwidth]{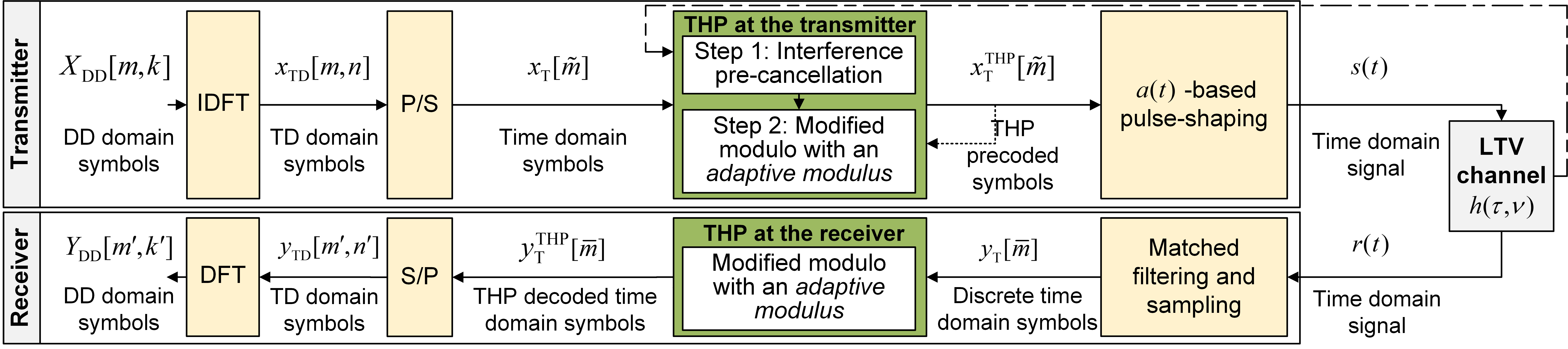}
  \caption{{Illustration} of the ODDM system with time domain THP, with the standalone ODDM system operations marked in yellow and the proposed THP operations marked in green.}\label{FIG3}
  \vspace{-1.5em}
\end{figure*}
\subsection{Transmitter}
At the transmitter, the time domain symbols $x_{\rm T}[\tilde m]$ obtained after IDFT and P/S conversion will be {sent} through THP operations. {The} THP at the transmitter involves two operations, namely, the element-wise successive ISI pre-cancellation and the modulo operation. The resulting time domain symbols after each of these operations are denoted by $x^{\rm{IC}}_{\rm{T}}[\tilde m]$ and $x^{\rm{THP}}_{\rm{T}}[\tilde m]$, respectively, (where ``IC'' stands for interference cancellation). $x^{\rm IC}_{\rm T}[\tilde m]$ is mathematically expressed as:
\begin{equation}
\begin{aligned}
& x^{\rm \!IC}_{\rm\! T}\![\tilde m] {=}\!\! \begin{cases}0 ,& \!\!\!\! \!\!\!-L_{\rm CP}\!\le \!{\tilde m}\!<\!0,\\
  x_{\rm \!T}\![\tilde m] \!\!-\!\!\!\! \sum\limits_{p=2}^{P}\!\!\!\!\frac{h_{p}\![\tilde m\!-\!l_{p}\!] \!{x}^{\rm \!THP}_{\rm T}\![\tilde m\!-\!l_{p}\!]}{h_1[\tilde m]}, & \!\! \!\!\!{0{\le} \tilde m {\le} M\!N\!\!-\!\!1},
\end{cases}
\end{aligned}
\label{THPEqu19B}
\end{equation}
where $h_p[\tilde m] = {h_p}{e^{j2\pi {\nu _p}\tilde m\frac{T}{M}}}$ refers to the LTV channel fading coefficient of the $p$th propagation path at $\tilde m \frac{T}{M}$, {\color{black}$\sum\limits_{p=2}^{P}\!\!\!\!\frac{h_p\![\tilde m\!-\!l_{p}\!] \!{x}^{\rm \!THP}_{\rm T}\![\tilde m\!-\!l_{p}\!]}{h_1[\tilde m]}$ denotes the ISI suffered by $x_{\rm \!T}\![\tilde m]$}. \par

We clarify that after the ISI pre-cancellation, energy of $x^{\rm{IC}}_{\text{T}}[m]$ may increase or {diverge}, especially when $MN$ is large and $|h_1[\tilde m]|$ is small \cite{b4}. To address this, the modulo operation is introduced in THP to constrain the power of the ISI pre-cancelled symbols \cite{b4}. Accordingly, the TH precoded signal after the modulo operation is obtained as:
\begin{figure}
  \centering
  \includegraphics[width=.45\textwidth]{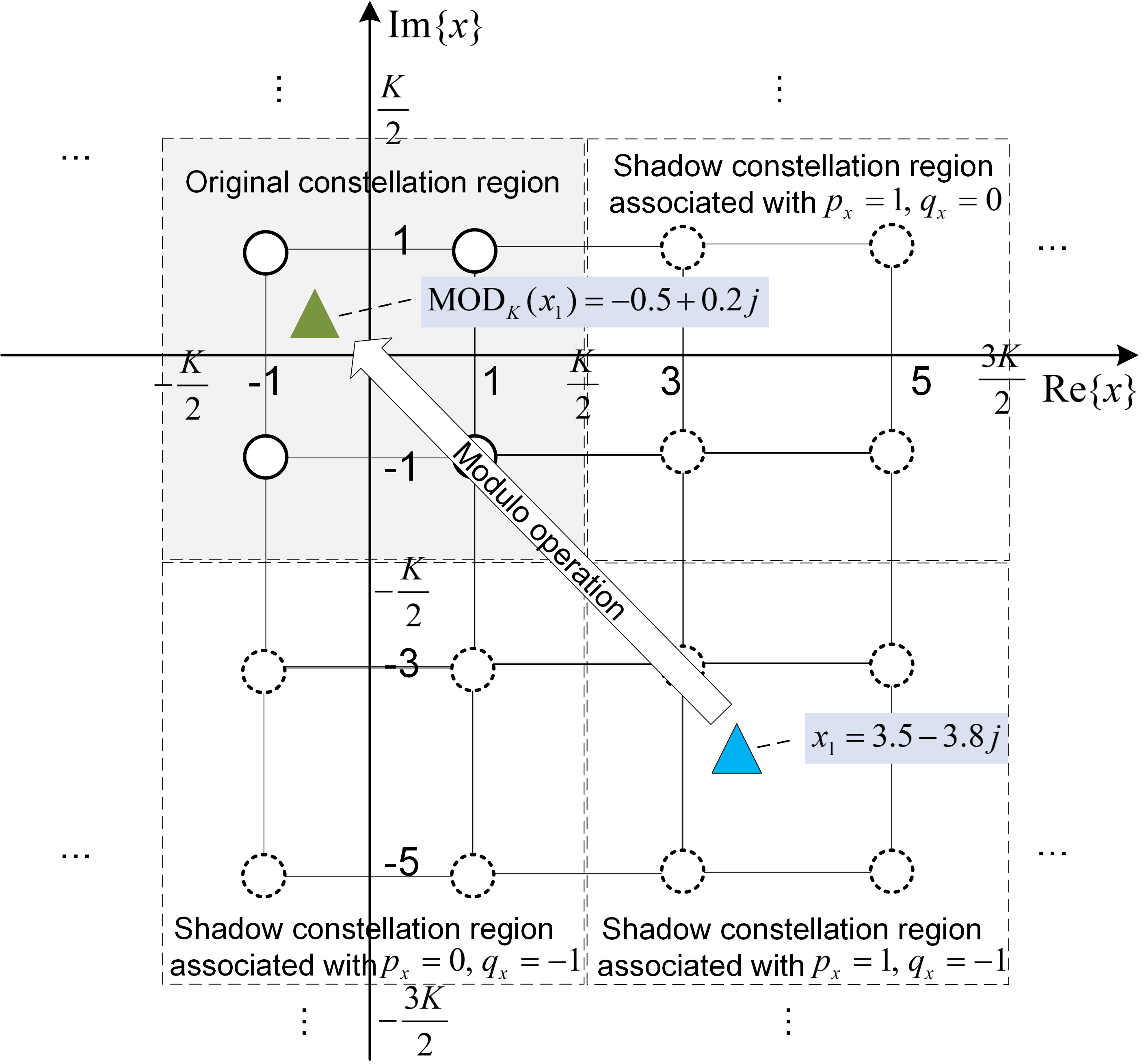}
  \caption{{\color{black}Visualization of the modulo operation in (\ref{THPEqu4}) for 4-QAM (i.e., $K = 4$). Herein, we consider the modulo operation on the symbol ($x_1 = 3.5-3.8j$) marked by a blue triangle. $x_1$ falls in the shadow constellation region corresponding to $p_x=1$ and $q_x=-1$. After modulo operation, its value becomes ${\rm MOD}_K(x_1)=-0.5+0.2j$ and the symbol falls back inside the original constellation region.  (The original constellation region (${\rm Re}\{x\}\in [-\frac{K}{2},\frac{K}{2}]$ and ${\rm Im}\{x\}\in [-\frac{K}{2},\frac{K}{2}]$) is shown in gray shading, and the three shadow constellation regions are shown in white shading.)}}
  \label{MODULOOPERATION}
\end{figure}

\begin{equation}
x^{\rm THP}_{\rm T}[\tilde m] =  {\rm MOD}_{K}\left(x^{\rm IC}_{\rm T}[\tilde m]\right),
\label{THPEqu19A}
\end{equation}
where ${\rm MOD}_{K}(x)$ denotes the modulo operation\footnote{We note that $[x]_{NM}$ in (\ref{Equ9BXXX}) and ${\rm MOD}_{K}(x)$ in (\ref{THPEqu19A}) are referred to as modulo operations in OTFS and THP literature, respectively \cite{bb4,b4}. However, we clarify that they differ from the domain to which they are defined, i.e., $[x]_{NM}$ is defined for real values of $x$ and ${\rm MOD}_{K}(x)$ is defined for complex values of $x$. Due to this, it can be considered that (\ref{THPEqu4}) is simply the extension of the real-valued modulo operation in (\ref{Equ9BXXX}) to a complex domain.} with the modulus $K$:
\begin{equation}
\begin{aligned}
{\rm MOD}_{K}(x) \!=\! x\!-\!\underbrace{\left\lfloor \frac{{\rm Re}\{x\}}{K} \!+ \!\frac{1}{2}\right\rfloor }_{p_{x}}\!K\!-j\underbrace{\left\lfloor \frac{{\rm Im}\{x\}}{K} \!+\! \frac{1}{2}\right\rfloor}_{q_x}\!K,
\end{aligned}
\label{THPEqu4}
\end{equation}
where $p_x$ and $q_x$ are integer multiples of ${\rm Re}\{x\}\! +\frac{K}{2}$ and ${\rm Im}\{x\}\! +\frac{K}{2}$ w.r.t. the modulus $K$, respectively, and $\lfloor \cdot \rfloor$ refers to the floor function. A visualization of the modulo operation in (\ref{THPEqu4}) is given in Fig. \ref{MODULOOPERATION}. We can observe from Fig. \ref{MODULOOPERATION} that, after the modulo operation, the obtained symbols are included in the original constellation region bounded by ${\rm Re}\{{\rm MOD}_{K}(x)\} \in [-\frac{K}{2}, \frac{K}{2}]$ and ${\rm Im}\{{\rm MOD}_{K}(x)\} \in [-\frac{K}{2}, \frac{K}{2}]$. Thus, the exceeding energy of the ISI pre-cancelled symbols is constrained.\par

{\color{black} Furthermore, if the real and imaginary components of the ISI term in (\ref{THPEqu19B}) are respectively large w.r.t. $\frac{K}{2}$, the real and imaginary components of $x^{\rm THP}_{\rm TD}[m',n']$ obtained after the modulo operation in (\ref{THPEqu19A}) can be considered to follow uniform distributions \cite{b1}, i.e., ${\rm Re}\{x^{\rm THP}_{\rm TD}[m',n']\} \sim {\cal U}(-\frac{K}{2},\frac{K}{2})$ and ${\rm Im}\{x^{\rm THP}_{\rm TD}[m',n']\} \sim {\cal U}(-\frac{K}{2},\frac{K}{2})$. Based on the above, the average power of the THP precoded symbols $x^{\rm THP}_{\rm TD}$, denoted as ${\cal E}_{x^{\rm THP}_{\rm TD}}$, can be obtained as:
\begin{equation}
\begin{aligned}
{\cal E}_{x^{\rm THP}_{\rm TD}} & = {{E}[x^{\rm THP}_{\rm TD}[m',n']^2]} & = 2\int^{\frac{K}{2}}_{-\frac{K}{2}} \frac{x^2}{K} dx= \frac{K^2}{6}.
\end{aligned}
\label{DEMOX8}
\end{equation}
Besides, the maximum energy of $x^{\rm THP}_{\rm TD}$ can be obtained as $\frac{K^2}{2}$. Therefore, the peak-to-average power ratio (PAPR) of the proposed design is:
\begin{equation}
\begin{aligned}
{\rm PAPR}_{\rm THP-ODDM} = 3.
\end{aligned}
\label{PAPR}
\end{equation}
\begin{remark}

  {\color{black} To understand the accuracy of the derived results in (\ref{PAPR}), and to further demonstrate the impact of the modulo operation in the proposed design, in Fig. \ref{FIGPAPR}, we examine the PAPR of the proposed design while comparing it with those of  i) the classic ODDM system and ii) the proposed design without the modulo operation. To this end, in Fig. \ref{FIGPAPR}, the complementary cumulative distribution functions (CCDFs) of the PAPR of the respective systems are plotted. Therein, $M = 512, N = 64, K = 4, P = 7$, and 4-QAM modulation is adopted. Besides, the multipath delays are set to be different, the maximum Doppler shift is obtained based on the Jakes model while using the maximum Doppler shift of 1000 Hz, and $h_p \sim {\cal CN}(0,\frac{1}{P})$ \cite{bb4}.\par

  First, we can observe that the PAPR of the proposed design could be lower than that of the classical ODDM system. This is because the modulo operation in (\ref{THPEqu19A}) restricts the peak value of time domain symbols, while no such restriction is introduced in the ODDM system with THP but without the modulo operation. Second, we observe that if the modulo operation is not adopted, the PAPR of the proposed design will increase sharply due to the energy accumulation effect caused by the serial interference pre-cancellation in (\ref{THPEqu19B}). In particular, the PAPR of the design without modulo operation could be higher than 20 dB with a probability of $70\%$. This shows that the modulo operation is a necessary component for our design, and in general for all designs with THP.\par}
    {\color{black}Third, we find that although we analytically derive the PAPR of our design is $10log_{10}3 = 4.77\ {\rm dB}$ in (\ref{PAPR}), the possibility that the simulated PAPR is less than or equal to 4.77 dB is 89\% instead of 100 \%. This difference in the analytical results and simulation results arises because, following the classic THP literature \cite{b4,b8,A3,blight,b1}, we assumed that the real and imaginary components of $x^{\rm THP}_{\rm TD}[m',n']$ are uniformly distributed. However, for certain scenarios where the ISI energy in (\ref{THPEqu19B}) is small, this assumption could be invalid, leading to the mismatch.}

\end{remark}
}
  \begin{figure}[t]
  \centering
  \includegraphics[width=.45\textwidth]{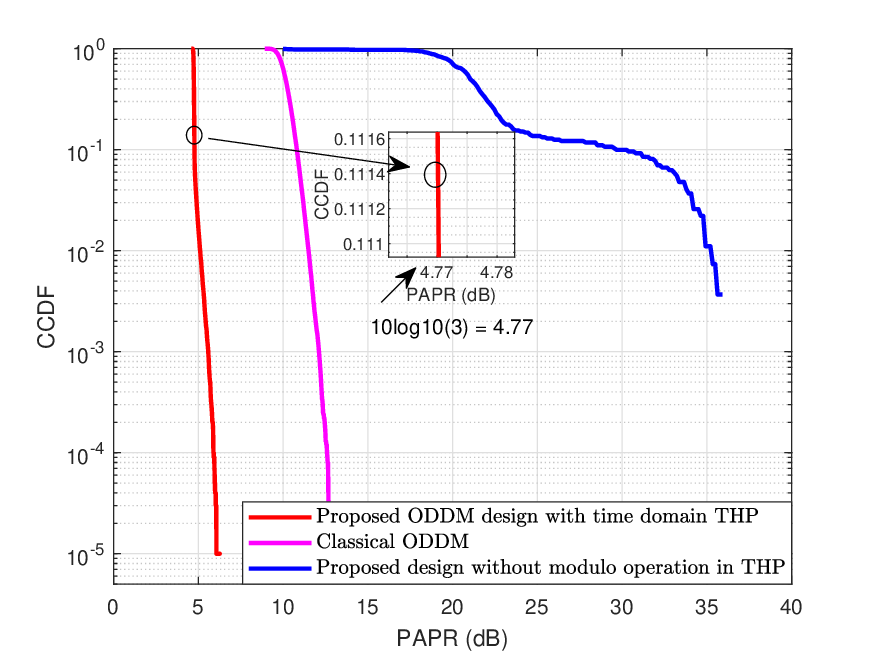}
  \caption{\color{black}{PAPR of the ODDM system with time domain THP.}}
  \label{FIGPAPR}
  \end{figure}
\begin{figure*}[b]
\setcounter{equation}{20}
\hrulefill
\begin{equation}
\begin{aligned}
\label{DEMOX4}
  y_{\mathrm{T}}^{\mathrm{THP}}[\bar{m}]&={\rm MOD} _{K_r}\left(\frac{h_1^*[\bar{m}]y_{\mathrm{T}}[\bar{m}]}{\left|h_1[\bar{m}]\right|} \right)\\
  &=|h_1| x_{\mathrm{T}}[\bar{m}]\underbrace{-\left\lfloor\frac{{\rm Re}\{|h_1|  x_{\mathrm{T}}[\bar{m}]+\tilde{w}[\bar{m}]\}}{K_r}+\frac{1}{2}\right\rfloor K_r-j\left\lfloor\frac{{\rm Im}\{|h_1| x_{\mathrm{T}}[\bar{m}]+\tilde{w}[\bar{m}]\}}{K_r}+\frac{1}{2}\right\rfloor  K_r}_{I_{\mathrm{T}}[\bar{m}]}+\tilde{w}[\bar{m}].
\end{aligned}
\end{equation}
\setcounter{equation}{18}
\end{figure*}
Afterwards, we clarify that the modulus $K$ in (\ref{THPEqu4}) needs to be designed according to the pdf of the symbols before THP \cite{b4}. In particular, $K$ needs to be selected such that the pdf of the constellation points in the original constellation region and the pdf of the constellation points in the shadow constellation region do to not overlap. To clearly understand this, we provide a visual demonstration in Fig. \ref{MODULONEW}, where we plot the pdfs of time domain symbols in (i) our system with 4-QAM modulation and (ii) the classical THP system with 4-QAM modulation, i.e., {\color{black}the single-carrier} system with time domain data mapping and time domain THP. First, in the classical THP system \cite{b4}, $K = 2\sqrt{\cal M}$ is adopted when $\cal M$-QAM modulation is used. This is due to the fact that in classical THP systems, the symbols before THP lie in the $\cal M$-QAM constellation points $\{a_{\rm R} + a_{\rm I}j | a_{\rm R}, a_{\rm I} \in {\pm 1, \pm 3,..., \pm \sqrt{\cal M} -1}\}$. Thus the pdf of QAM symbols is confined to the region $-\sqrt{\cal M}$ to $\sqrt{\cal M}$; hence, $K = 2\sqrt{\cal M}$. As shown in Fig. \ref{MODULONEW}(b), for classical THP systems with 4-QAM, $K=2\sqrt{4}=4$ is adopted. 

 However, in our design, $\cal M$-QAM symbols are embedded in the DD domain and the symbols before THP are obtained by performing DFT on the $\cal M$-QAM DD domain symbols. Due to this, the symbols before THP do not lie on the $\cal M$-QAM constellation. As an example shown by Fig. \ref{MODULONEW}(a), for 4-QAM DD domain symbols, the pdf of the symbols before THP do not lie within the region bounded by the $-\sqrt{\cal M}$ to $\sqrt{\cal M}$, i.e., -2 to +2. Due to this, the modulus $K=2{\sqrt{\cal M}}$ used in classical THP systems may not be suitable for our system. As a solution, we consider a {\it modified} modulus $K=2\alpha{\sqrt{\cal M}}$ with {a scaling factor $\alpha$}. The selection of suitable values for $\alpha$ will be discussed in detail in Sections IV and V.\par

\begin{figure}[t]
  \centering
  \includegraphics[width=.45\textwidth]{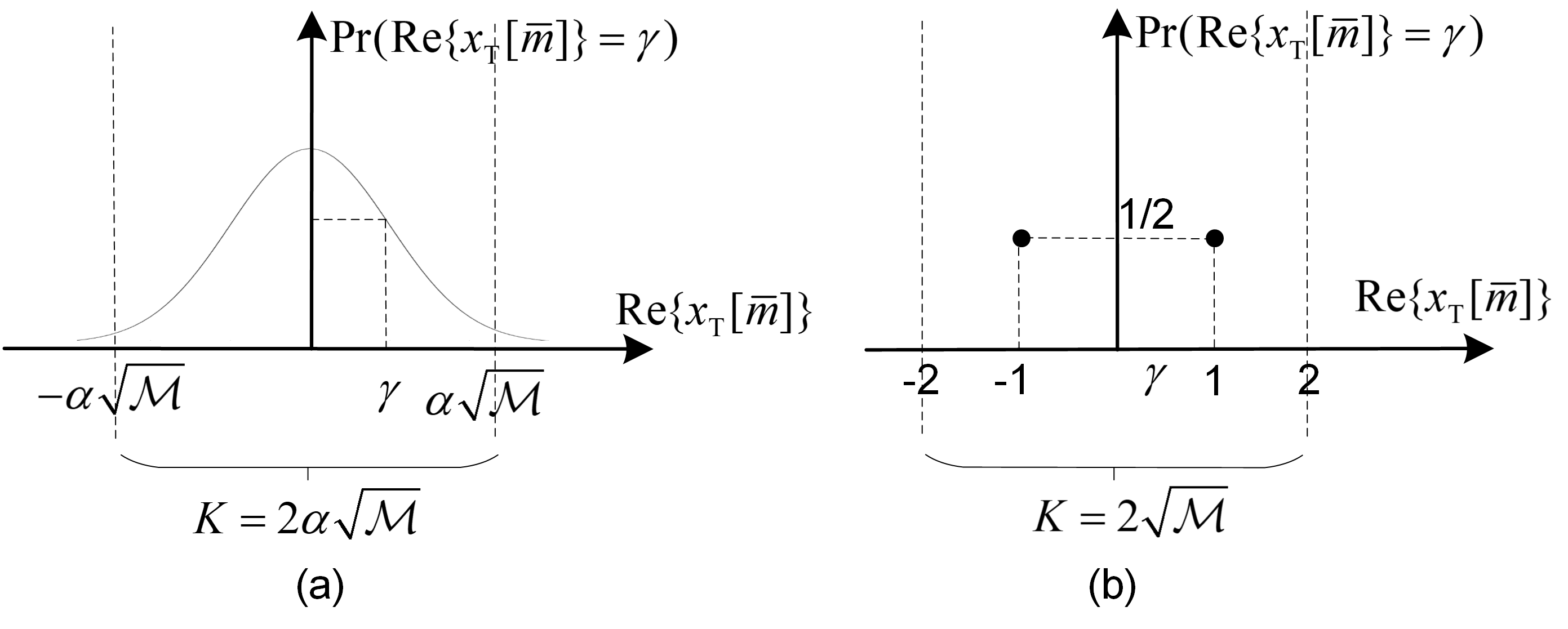}
  \caption{Pdf of the symbols before THP for (a) the proposed design and (b) classical THP system. For simplicity, 4-QAM modulation is considered, i.e., ${\cal M} = 4$. Only the pdfs of signals' real components are demonstrated and those of imaginary components are same. }
  \label{MODULONEW}
  \vspace{-1.5em}
\end{figure}
Finally, we note that after THP operations at the transmitter, $x^{\rm THP}_{\rm T}[\tilde m]$ is convoluted with the SRN pulse for transmission, which is similar to that in (\ref{Equ3}).\par
\begin{remark}
  Now we can observe the noteworthy distinctions between the proposed ODDM system with time domain THP and the classical THP systems \cite{b4}. Firstly, while the classical THP systems operates under the LTI channel, our proposed design accommodates the LTV channel information. Secondly, regarding the modulo operation, classical THP systems employ the modulus $K = 2\sqrt{\mathcal{M}}$, whereas our design adopts the modulus $K = 2\alpha\sqrt{\mathcal{M}}$, reflecting the consideration that data multiplexing and THP in distinct domains.
\end{remark}

\begin{remark}
Despite considering time domain THP for the ODDM system in this work, we clarify that DD domain THP is still possible for the ODDM system through two approaches {\color{black}based on the perfect CSI assumption}. Firstly, ${\bf H}_{\rm DD}$ in (\ref{Equ13}) can be QR decomposed and then precoding can be made feasible based on the resulting triangle matrix \cite{A3}. Secondly, as proposed in \cite{bb15,b9,ba}, null symbols can be inserted to $X_{\rm DD}[m,k]$, similar to the way that zeros are included in the zero-padded OTFS for enabling DFE equalization. Despite these possibilities, it is worth noting that QR decomposing ${\bf H}_{\rm{DD}}$ would lead to a high computational complexity of ${\cal O}(N^3M^3)$, and inserting null symbols into $X_{\rm{DD}}[m,k]$ {\color{black}for THP could} compromise resource efficiency. {\color{black}Consequently, the proposed ODDM system with time domain THP could realize lower precoding complexity and higher resource efficiency.
}\par
\end{remark}

\subsection{Receiver}

After matched filtering, sampling, as shown in Fig. \ref{FIG3}, and discarding the zero prefix in (\ref{THPEqu19B}), the received discrete time domain symbols are obtained as:
\begin{equation}
\begin{aligned}
& y_{\rm T}[\bar m] = \sum^P_{p=1}{h_{p}[\bar m-l_{p}]}x^{\rm THP}_{\rm T}[{\bar m}-l_{p}]+ w[\bar m],
 \end{aligned}
\label{DEMO0}
\end{equation}
where $ \bar m \in \{0,1,...,NM-1\}$. When carefully using (\ref{THPEqu19B})-(\ref{THPEqu4}) in (\ref{DEMO0}), the $y_{\rm T}[\bar m]$ can be given in terms of $x_{\rm T}[{\bar m}]$ as:
\begin{equation}
\begin{aligned}
y_{\rm T}[{\bar m}]= {h_1}[\bar m] x_{\rm T}[{\bar m}]{-}h_1[\bar m](p_{\bar m}K {+}q_{\bar m}Kj){+} w[\bar m],
\end{aligned}
\label{DEMOX3}
\end{equation}
where $p_{\bar m}$ and $q_{\bar m}$ are integers and $w[\bar m] \sim {\cal CN}(0,\sigma^2_w)$.\par

Thereafter, to eliminate $p_{\bar m}K$ and $q_{\bar m}Kj$ in (\ref{DEMOX3}), the receiver THP operation is performed on $y_{\rm T}[{\bar m}]$, as demonstrated in Fig. \ref{FIG3}. In particular, the modulo operation is performed on $\frac{h^*_1[\bar m]}{|h_1[\bar m]|}y_{\rm T}[{\bar m}]$ with the modulus $K_r = |h_1|K$ following (\ref{THPEqu19A}). In doing so, the {THP decoded} time domain signal $y_{\mathrm{T}}^{\mathrm{THP}}[\bar{m}]$ is obtained as (\ref{DEMOX4}), which is given at the bottom of this page. Therein, ${I_{\rm T}[\bar m]}$ is the time domain interference generated by the modulo operation in THP, ${\tilde {w}}[\bar m]=\frac{h^*_1[\bar m]}{|h_1[\bar m]|}w[\bar m]$, and ${\tilde {w}}[\bar m] \sim {\cal {CN}}(0,\sigma^2_w)$.\par

Subsequently, similar to that in the standalone ODDM system, $y^{\rm THP}_{\rm T}[\bar m]$ goes through the S/P converter to obtain $y^{\rm THP}_{\rm TD}[m',n']$ with $y^{\rm THP}_{\rm T}[\bar m] = y^{\rm THP}_{\rm T}[n'M+m'] = y^{\rm THP}_{\rm TD}[m',n']$, $m'\in \{0,1,...,M-1\}$, and $n' \in \{0,1,...,N-1\}$. Finally, $y^{\rm THP}_{\rm TD}$ is transformed into DD domain to obtain $Y_{\rm DD}[m',k']$ based on (\ref{Equ2}) and (\ref{Equ12}) as:
\setcounter{equation}{21}
\begin{equation}
\begin{aligned}
\!\!\!Y_{\rm DD}[m',k']{=} |h_1|\!X_{\rm DD} [m',k']{+}I_{\rm DD}[m',k']{+} {w}[m',k'].
\end{aligned}
\label{DEMOX5}
\end{equation}
Therein, we refer to $I_{\rm DD}[m',k']$ as the {\it modified modulus interference}, which is given by:
\begin{equation}
\begin{aligned}
I_{\rm DD}[m',k'] = \frac{1}{\sqrt{N}}\sum^{N-1}_{n'=0}I_{\rm TD}[m',n']e^{-j2\pi\frac{n'k'}{N}},
\end{aligned}
\label{DEMOX6}
\end{equation}
where $I_{\rm TD}[m',n'] = I_{\rm T}[n'M+m'] = I_{\rm T}[\bar m]$.\par

When carefully observing the $Y_{\rm DD}[m',k']$ in (\ref{DEMOX5}), while comparing the $Y_{\rm DD}[m',k']$ in (\ref{Equ12}), we find that the introduction of the time domain THP to the ODDM system makes the $Y_{\rm DD}[m',k']$ immune to ISI. Due to this, it becomes possible to perform DD domain single-tap equalization for the ODDM system with time domain THP \footnote{\color{black}We note that the proposed THP design or other variants of THP designs can also be
introduced to reduce the receiver complexity other systems suffering from severe ISI, e.g., the system adopting the unified non-orthogonal waveform which introduces ISI at the transmitter to improve the system spectrum efficiency \cite{unow}. Besides, given the similarity in the digital sequence of ODDM and OTFS, and the fact that our THP designs rely on the time-domain digital symbols to perform THP, we clarify that the proposed THP design can readily be applied to OTFS systems.}. The single-tap equalized DD domain symbols can be obtained as:
\begin{equation}
\begin{aligned}
{\hat Y}_{\rm DD}[m',k'] & = \frac{{Y}_{\rm DD}[m',k']}{|h_1|}\\
& = X_{\rm DD} [m',k']{+}\frac{I_{\rm DD}[m',k']}{|h_1|}{+}\frac{{w}[m',k']}{|h_1|}.
\end{aligned}
\label{DEMOST}
\end{equation}

As can be observed in (\ref{DEMOST}), in addition to the effects of {the first propagation path} (i.e., $h_1$) and noise (i.e., ${w}[m',k']$), the system reliability is influenced by the modified modulus interference $I_{\rm DD}[m',k']$. This interference term {needs to be} as small as possible to ensure reliable communication. Hence, in the next section, we analyze the modified modulus interference in detail.
\section{BER Performance Analysis}
In this section, we analytically investigate the BER performance of the proposed design. We note that the non-linear modulo operations in THP make it difficult to directly analyze $I_{\rm DD}[m',k']$ and ${{w}[m',k']}$ by obtaining their pdfs. Thus, we decompose the impacts of $I_{\rm DD}[m',k']$ and ${{w}[m',k']}$ on BER into three {parts}, namely {\it power loss}, {\it modulo noise loss}, {\it modulo signal loss}, and then analyze them one after the other{\footnote{Shannon's capacity formula for AWGN channels dictates that the capacity-achieving input should be Gaussian variables, a criterion that TH precoded symbols do not satisfy. Due to this, systems with THP encounter another type of loss, referred to as shaping loss \cite{b1}. The shaping loss affects the channel capacity of our system, and it will be considered in future works.}. Based on this, we obtain the lower bound of the BER for the ODDM system with time domain THP. \par

Prior to the analysis, we denote the BERs associated with the {\it power loss}, {\it modulo noise loss}, {\it modulo signal loss} for the ODDM system with time domain THP as ${\cal P}^{\rm PL}_{e}$, ${\cal P}^{\rm MNL}_{e}$, and ${\cal P}^{\rm MSL}_{e}$ respectively. Also, we denote the average energy of the $\cal M$-QAM modulated symbols $X_{\rm DD} [m,k]$ as ${{\cal E}_{X_{\rm DD}}}$. When ${\rm log}_2{\cal M}$ is even and the minimum distance between {\color{black}constellation points} is two, the average energy of ${\cal E}_{X_{\rm DD}}$ becomes:
\begin{equation}
{\cal E}_{X_{\rm DD}} = \frac{2({\cal M}-1) }{3}.
\label{DEMOX9}
\end{equation}

\subsection{Performance Analysis for ODDM with 4-QAM Modulation}
\label{SECTIONIVA}
In this subsection, we analyze the BERs of our ODDM system with time domain THP for 4-QAM modulation by characterizing the {\it power loss}, {\it modulo noise loss}, {\it modulo signal loss}. The BER associated with 16-QAM modulation will be analyzed later in Section \ref{SECTIONIVB}.

\subsubsection{Power Loss}
According to (\ref{DEMOX6}) and the expression ${I_{\mathrm{T}}[\bar{m}]}$ in (\ref{DEMOX4}), we note that ${I_{\rm DD}[m',k']}$ can be ignored if $K$ is significantly large compared to the real and imaginary components of $x_{\mathrm{T}}[\bar{m}]+\tilde{w}[\bar{m}]/|h_1|$. Then (\ref{DEMOST}) can be simplified into:
\begin{equation}
\label{IGNOREI}
  {\hat Y}_{\rm DD}[m',k'] {\approx} X_{\rm DD} [m',k']{+}\frac{{w}[m',k']}{|h_1|},
\end{equation}
which can be regarded as the signal obtained after single-tap equalized signal in {an} ISI-free system (i.e., single-tap system) that does not adopt THP. In this case, the system BER is only affected by the AWGN and the channel coefficient, and is easy to be obtained.\par
However, we specify that it is not viable for our design to set an arbitrarily large $K$ to eliminate ${I_{\rm DD}[m',k']}$. This limitation arises from the fact that the power of TH precoded time domain symbols escalates with $K$, surpassing the energy of the unprecoded DD domain symbols \cite{b1}. This BER performance degradation caused by the increased power of the TH precoded symbols is called {\it power loss} \cite{b1}. Next, we present the BER associated with the {\it power loss}, denoted as ${\cal P}^{\rm PL}_{e}$.\par


By comparing (\ref{DEMOX8}) and (\ref{DEMOX9}), it is evident that for large values of $\alpha$, the average power of the THP precoded symbols ${\cal E}_{x^{\rm THP}_{\rm TD}}$ becomes more significant than that of the unprecoded symbols ${\cal E}_{X_{\rm DD}}$. Following (\ref{DEMOX9}), (\ref{DEMOX8}), and defining the SNR as $\Omega = \frac{{\cal E}_{x^{\rm THP}_{\rm TD}}}{\sigma^2_{w}}$, the noise variance $\sigma^2_{w}$ can be expressed as:
\begin{equation}
\begin{aligned}
\sigma^2_{w} & = \frac{\alpha^2{\cal E}_{X_{\rm DD}}+\frac{2}{3}\alpha^2}{\Omega}.
\end{aligned}
\label{DEMOX88}
\end{equation}
Based on the formula of $\sigma^2_{w}$ above, we can derive the BER for the system defined by (\ref{DEMOST}). It is important to note that if $I_{\rm DD}[m',k']$ is disregarded, the BER related to the {\it power loss} of (\ref{IGNOREI}) is lower compared to the BER for the system defined by (\ref{DEMOST}). Therefore, a lower bound of ${\cal P}^{\rm PL}_{e}$ for the proposed design with 4-QAM can be obtained as the BER for the system defined by (\ref{IGNOREI}). The result is given in the following proposition.\par
\begin{prop}
BER caused by the power loss with 4-QAM modulation is lower bounded by:
\begin{subequations}
\begin{equation}
{\begin{aligned}
 &{\cal P}^{\rm PL}_{e,\ {\rm 4-QAM}} \ge \frac{1}{2}\left(1-\sqrt{\frac{\frac{{{\cal E}_{X_{\rm DD}}\sigma^2_{h_1}}}{\sigma_w^2}}{2+\frac{{{\cal E}_{X_{\rm DD}}\sigma^2_{h_1}}}{\sigma_w^2}}}\right)
\end{aligned}}
\label{DEMOX10a}
\end{equation}
\begin{equation}
{\begin{aligned}
= \!\frac{1}{2}\!\left(\! {1 \!- \!\sqrt {\frac{{\frac{3{\sigma _{{h_1}}^2\Omega}}{{4{\alpha ^2}}}}}{{2 + \frac{3{\sigma _{{h_1}}^2\Omega}}{{4{\alpha ^2}}}}}} } \!\right)
\end{aligned}}
\label{DEMOX10b}
\end{equation}
\begin{equation}
{\begin{aligned}
=\! \frac{{2{\alpha ^2}}}{{3\sigma^2_{h_1}\Omega}} + o\left({{\frac{{4{\alpha ^2}}}{3{\sigma _{{h_1}}^2\Omega}}}}\right),
\end{aligned}}
\label{DEMOX10}
\end{equation}
\end{subequations}
{where} $\alpha$ is sufficiently large, $h_1 \sim {\cal{CN}}(0,\sigma^2_{h_1})$, and ${\rm Re}\{x^{\rm THP}_{\rm TD}[m,n]\},{\rm Im}\{x^{\rm THP}_{\rm TD}[m,n]\} {\sim} {\cal U}(-\alpha\sqrt{\cal M},\alpha\sqrt{\cal M})$.
\end{prop}
The right-hand side (RHS) of (\ref{DEMOX10a}) represents the BER for the system defined by (\ref{IGNOREI}). It is derived using (3.19) in \cite{b2} for 4-QAM symbols as:
\begin{equation}
\begin{aligned}
E_{h_1}\left[{Q\left( {|{h_1}|\frac{a}{{\sqrt {\frac{{\sigma _w^2}}{2}} }}} \right)}\right] = \frac{1}{2}\left(1-\sqrt{\frac{\sigma^2_{h_1}a^2}{\sigma_w^2+\sigma^2_{h_1}a^2}}\right),
\end{aligned}
\label{Equ43}
\end{equation}
where $Q(\cdot)$ is the tail distribution function of the standard normal distribution and $h_1\sim{\mathcal {CN}}(0,\sigma^2_{h_1})$ is assumed, where $\sigma^2_{h_1}$ is the variance of $h_1$. In addition, (\ref{DEMOX10b}) is obtained based on $ {{\cal E}_{X_{\rm DD}}}= \frac{3}{4\alpha^2}{{\cal E}_{x^{\rm THP}_{\rm TD}}}$, (\ref{DEMOX8}), and (\ref{DEMOX9}). Moreover, (\ref{DEMOX10}) is obtained by using Taylor expansion on (\ref{DEMOX10b}). \par

We can observe from (\ref{DEMOX10}) that when $\sigma_{h_1}$ is given, ${\cal P}^{\rm PL}_{e}$ will increase quadratically with $\alpha$ and decrease inversely proportional to SNR. {\it Therefore, setting $\alpha$ arbitrarily large for the ODDM system with time domain THP is not feasible.}\par

\subsubsection{Modulo Noise Loss}
For the ISI-free system without THP, as illustrated by (\ref{IGNOREI}), it is evident that it remains unaffected by the modulo operation. However, in systems employing THP, such as the one depicted in Fig. \ref{MODULOOPERATION}, the impact of noise could become more severe after the modulo operation is performed at the receiver. Specifically, symbols received at the constellation boundary may be mistaken for symbols at the opposite boundary due to noise \cite{b1}. This BER performance degradation resulting from noise, which is characterized by the modulo operation, is referred in the THP literature as the {\it modulo noise loss} \cite{b1}.\par

We note that it is impossible to directly analyze the impact of the {\it modulo noise loss} on $I_{\mathrm{T}}[\bar{m}]$. This is because $x_{\mathrm{T}}[\bar m]$s do not reside on constellation points, instead they are randomly located as shown by Fig. \ref{MODULONEW}. However, we note that the BER due to the {\it modulo noise loss} of our system with time domain THP, ${\it P}_e^{\rm MNL}$, could be more severe that of the classical THP system. This is because performing the modulo operation on $|h_1| x_{\mathrm{T}}[\bar{m}]+\tilde{w}[\bar{m}]$ in (\ref{DEMOX4}) complicates the acquisition of the $\cal M$-QAM DD domain symbols in our design, whereas the $\cal M$-QAM symbols are initially mapped in the classical THP system. For the classical THP systems, the signal obtained after single-tap equalization can be written as:
\begin{equation}
\begin{aligned}
&{\hat Y}_{\rm DD}[m',k']\\
&  = X_{\rm DD} [m',k'] -K\\
& \times\!\! {{\left\lfloor \frac{\!|h_1|X_{\rm DD} [m',k']{+} \!{\tilde w}[m',k'] {+} \!|h_1|\frac{K}{2} {+} j|h_1|\frac{K}{2}}{|h_1|K}\right\rfloor }}\!{+}\!\frac{{\tilde w}[m',k']}{|h_1|}.
\end{aligned}
\label{DEMOX16}
\end{equation}
Next, we derive ${\cal P}_e^{\rm MNL}$ using the conditional pdfs of $X_{\rm DD} [m',k']{+}\frac{{w}[m',k']}{|h_1|}$ in (\ref{DEMOX16}).\par

As an example, Fig. \ref{FIG4} provides the pdfs of signals' real components in different systems, where ${\rm Re}\{X_{\rm DD}[m', k']\} {=} 1$ and $|h_1| {=} 1$ are considered. Fig. \ref{FIG4}(a) shows the pdf of ${\rm Re}\{X_{\rm DD} [m',k']{+}\frac{{w}[m',k']}{|h_1|}\}$ for the single-tap system that does not adopt THP, characterized by (\ref{IGNOREI}). We can observe that the value regions resulting in bit errors, highlighted in red, are characterized by ${\rm Re}\{X_{\rm DD} [m',k']{+}\frac{{w}[m',k']}{|h_1|}\} \le 0$, which aligns with the classical BER analysis for $\cal M$-QAM symbols. In Fig. \ref{FIG4}(b), the pdf of ${\rm Re}\{|h_1|X_{\rm DD} [m',k']{+} {\tilde w}[m',k']\}$ for the system defined by (\ref{DEMOX16}) is shown, with the values that result in bit errors due to the {\it modulo noise loss} highlighted in red. In detail, Fig. \ref{FIG4}(c) provides an example illustrating how the {\it modulo noise loss} manifests in Fig. \ref{FIG4}(b). In Fig. \ref{FIG4}(c), we notice that when $\frac{K}{2}\le {\rm Re}\{X_{\rm DD} [m',k']{+} \frac{{\tilde w}[m',k']}{|h_1|}\} \le {K}$, the equalized signal ${\rm Re}\{{\hat Y}_{\rm DD}[m',k']\}$ falls within $-\frac{K}{2}\le{\rm Re}\{{\hat Y}_{\rm DD}[m',k']\}\le 0$ due to the modulo operation at the receiver (corresponding to the second term of (\ref{DEMOX16})). Consequently, ${\rm Re}\{X_{\rm DD} [m',k']\}$ will be determined as $-1$ when ${\rm Re}\{X_{\rm DD} [m',k'] = 1$ {was} transmitted, since the Hamming distance between ${\rm Re}\{{\hat Y}_{\rm DD}[m',k']\}$ and ${\rm Re}\{X_{\rm DD} [m',k']\} = -1$ is minimal, resulting in the {\it modulo noise loss}.\par

After carefully observing Fig. \ref{FIG4}, {we} found that the error regions associated with the {\it modulo noise loss} for the ODDM system with time domain THP and $\cal M$-QAM modulation are intermittent, which is different from those of the single-tap system with QAM modulation and no THP. Second, we note that the error regions in the pdf are affected by the setting of $K$. Based on these understandings, it can be concluded that the bit errors due to the {\it modulo noise loss} for 4-QAM occurs when ${\rm Re}\left\{{\tilde w}[m',k']\right\}$ satisfies:
\begin{figure}[t]
  \centering
  \includegraphics[width=.45\textwidth]{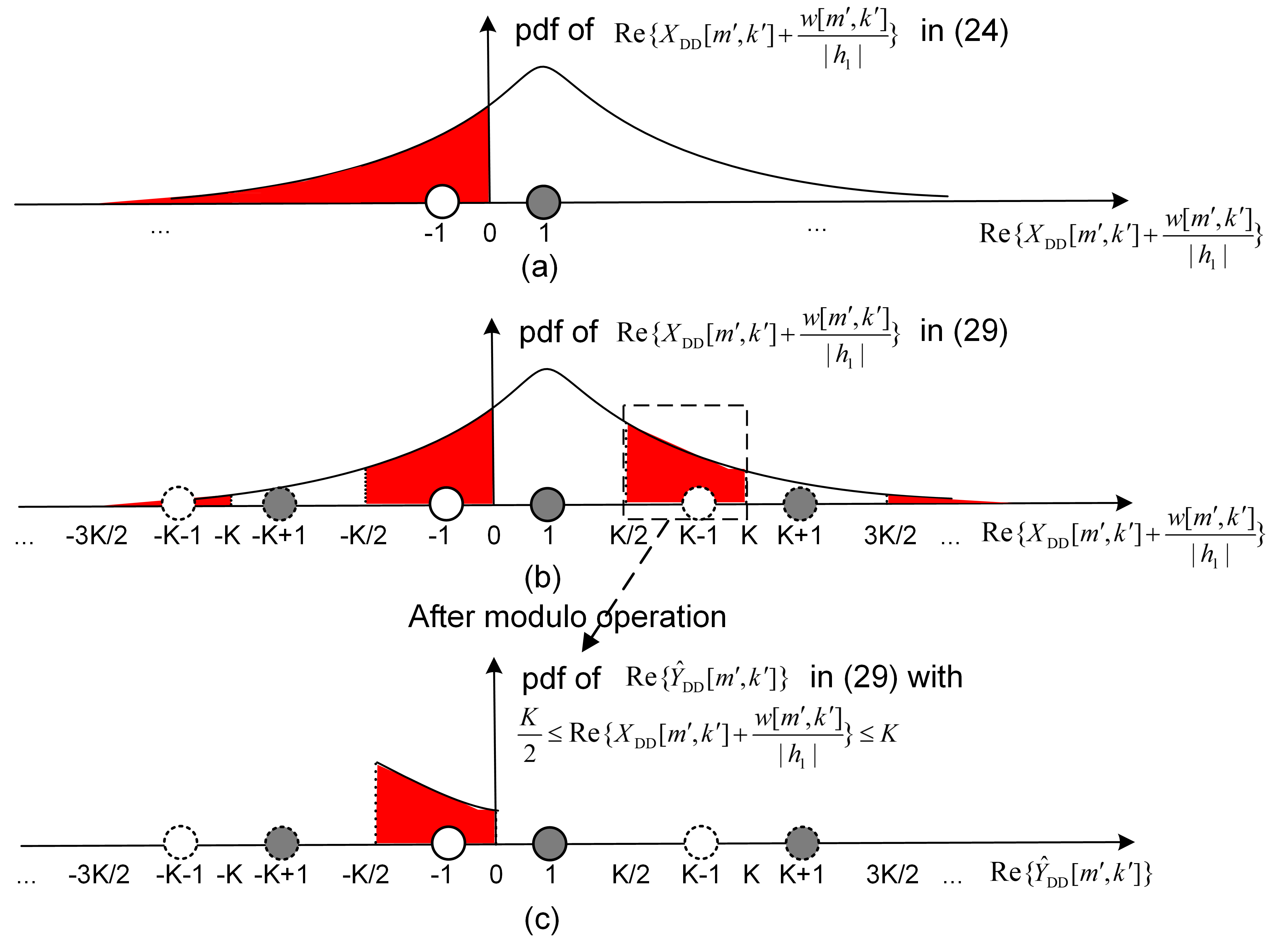}
  \caption{Illustration of values of signals' real components those lead to {\it modulo noise loss} when ${\rm Re}\{{X_{\rm DD} [m',k']} \}{=} 1$ and $|h_1| {=} 1$: (a) pdf of {\rm Re}\{$X_{\rm DD} [m',k']{+}\frac{{w}[m',k']}{|h_1|}$\} for the single-tap system that does not adopt THP, characterized by (\ref{IGNOREI}); (b) pdf of ${\rm Re}\{X_{\rm DD} [m',k']{+}\frac{{w}[m',k']}{|h_1|}\}$ for the proposed design, characterized by (\ref{DEMOX16}); (c) pdf of ${\rm Re}\{{\hat Y}_{\rm DD}[m',k']\}$ for the proposed design in (\ref{DEMOX16}) with $\frac{K}{2}\le {\rm Re}\{X_{\rm DD} [m',k']{+}\frac{{\tilde{w}[m',k']}}{|h_1|}\} \le {K}$. Signal values that result in bit errors are shaded in red.}\label{FIG4}
  \vspace{-1.5em}
\end{figure}
\begin{equation}
\begin{aligned}
{\rm Re}\left\{{\tilde w}[m',k']\right\} & \in \left\{|h_1|(\eta-\frac{1}{2}){K}-1\le {\rm Re}\left\{{\tilde w}[m',k']\right\}\right.\\
& \left.\le |h_1|\left 
(\eta{K}-1\right)\right\},
\end{aligned}
\end{equation} 
where $\eta$ is an integer. Since ${\tilde w}[m',k']$ is an isotropic complex Gaussian variable, i.e., the imaginary component of ${\tilde w}[m',k']$ follows the same distribution as its real component, we can obtain a lower bound of the BER due to the {\it modulo noise loss} for our design with 4-QAM as:
\begin{equation}
\begin{aligned}
& {\cal P}_{e,\ {\rm 4-QAM}}^{\rm MNL} \ge  E_{h_1}[\sum_{\eta}{\cal F}({( {\eta  - \frac{1}{2}} )K - 1},{\eta K - 1})],
\end{aligned}
\label{Equ42a}
\end{equation}
where ${\cal F}\left(a,b\right)$ is a positive number defined as:
\begin{equation}
\begin{aligned}
& {\cal F}\left(a,b\right) = \left|{Q\left( {|{h_1}|\frac{a}{{\sqrt {\frac{{\sigma _w^2}}{2}} }}} \right)}-Q{\left( {|{h_1}|\frac{b}{{\sqrt {\frac{{\sigma _w^2}}{2}} }}} \right)}\right|.
\label{Equ42}
\end{aligned}
\end{equation}

By carefully observing Fig. \ref{FIG4}, we can observe that ${\cal P}_{e,\ {\rm 4-QAM}}^{\rm MNL}$ is dominated by the case that $\eta= 0$ and $\eta = 1$. Based on this, a BER lower bound associated with the {\it modulo signal loss} for our design with 4-QAM modulations can be obtained, and it is given in the following proposition.\par
\begin{prop}
A lower bound for the BER due to the {\it modulo noise loss} with the 4-QAM modulation is:
\begin{equation}
\begin{split}
&  {\cal P}_{e,\ {4}-{\rm QAM}}^{\rm MNL} {\ge} \frac{1}{2}\sqrt {\frac{{\sigma _{{h_1}}^2{{\left( {4\alpha  {-} 1} \right)}^2}}}{{\sigma _w^2 {+} \sigma _{{h_1}}^2{{\left( {4\alpha  {-} 1} \right)}^2}}}}  {-} \frac{1}{2}\sqrt {\frac{{\sigma _{{h_1}}^2{{\left( {2 \alpha  {-} 1} \right)}^2}}}{{\sigma _w^2 {+} \sigma _{{h_1}}^2 {{\left( {2 \alpha  {-} 1} \right)}^2}}}}\\
& +\frac{1}{2}\sqrt {\frac{{\sigma _{{h_1}}^2{{\left( {2 \alpha + 1 } \right)}^2}}}{{\sigma _w^2 + \sigma _{{h_1}}^2{{\left( {2 \alpha + 1 } \right)}^2}}}} -  \frac{1}{2}\sqrt {\frac{{\sigma _{{h_1}}^2}}{{\sigma _w^2 + \sigma _{{h_1}}^2}}},
\end{split}
\label{DEMOX36}
\end{equation}
where $h_1\sim {\cal CN}(0,\sigma^2_{h_1})$ and $w[m,k]\sim {\cal CN}(0,\sigma^2_w)$.\par
\end{prop}

According to (\ref{DEMOX36}), we find that ${\cal P}_{e,\ {\rm 4-QAM}}^{\rm MNL}$ is jointly affected by $\sigma^2_w$, $\sigma^2_{h_1}$, and $\alpha$. Moreover, when $\alpha$ is sufficiently large, (\ref{DEMOX36}) can be simplified into:
\begin{equation}
\begin{aligned}
& {\cal P}_{e,\ {\rm 4-QAM}}^{\rm MNL} \ge \frac{1}{2}-  \frac{1}{2}\sqrt {\frac{{\sigma _{{h_1}}^2}}{{\sigma _w^2 + \sigma _{{h_1}}^2}}}.
\end{aligned}
\label{DEMOX37}
\end{equation}

We recognize that through straightforward derivations, it can be shown that the RHS of (\ref{DEMOX37}) is equivalent to the RHS of (\ref{DEMOX10a}). That is because, as $\alpha$ increases, the bit errors induced by the modulo operation decrease, and the system's BER approaches the BER of the single-tap system that does not adopt THP, as characterized by (\ref{IGNOREI}).\par
\subsubsection{Modulo Signal Loss}
As highlighted in Fig. \ref{MODULONEW} and its corresponding descriptions, the time domain symbols before THP operations at the transmitter, i.e., $x_{\rm T}[\bar m]$s, do not lie on the $\cal M$-QAM constellation points. This is because they are obtained by performing DFT on the $\cal M$-QAM symbols, as shown by Fig. \ref{FIG3}. Consequently, performing the modulo operation on $x_{\rm T}[\bar m]$ with a relatively small $K$ will introduce interference terms, as shown in (\ref{DEMOX4}), thereby resulting in additional performance loss. This type of loss is unique to our proposed design and does not occur in the classical THP systems \cite{b1}. We define the loss as the {\it modulo signal loss}.\par

{To examine} the {\it modulo signal loss}, we analyze (\ref{DEMOX4}) in the high SNR region while neglecting the noise term $\tilde{w}[\bar{m}]$. The term $I_{\rm TD}[m',n']$ (equals to $I_{\rm T}[n'M+m']$ in (\ref{DEMOX4})) can be denoted as $I^{\rm High\ SNR}_{\rm TD}[m',n']$:
\begin{equation}
\begin{aligned}
I^{\rm High\ SNR}_{\rm TD}[m',n'] = & -\left\lfloor\frac{{\rm Re}\{|h_1|  x_{\mathrm{TD}}[m',n']\}}{K_r}+\frac{1}{2}\right\rfloor K_r\\
& -j\left\lfloor\frac{{\rm Im}\{|h_1| x_{\mathrm{TD}}[m',n']\}}{K_r}+\frac{1}{2}\right\rfloor K_r.
\end{aligned}
\label{DEMOX11}
\end{equation}
Without loss of generality, we first investigate the value of the ${\rm Re}\{I^{\rm High\ SNR}_{\rm TD}[m',n']/|h_1|\}$, i.e., $-\left\lfloor\frac{{\rm Re}\{x_{\mathrm{TD}}[m',n']\}}{K}+\frac{1}{2}\right\rfloor K$. By traversing the values of ${\rm Re}\{x_{\rm TD}[m',n']\}$, it can be obtained:
\begin{equation}
\begin{aligned}
 {\rm Re}& \left\{{I^{\rm High\ SNR}_{\rm TD}[m',n']}/{|h_1|}\right\}= \\
& \begin{cases}
    0, & \mbox{if} -\frac{K}{2}\le{\rm Re}\{x_{\rm TD}[m',n']\}<\frac{K}{2}\\
    -K, & \mbox{if } \frac{K}{2}\le{\rm Re}\{x_{\rm TD}[m',n']\}<\frac{3K}{2} \\
    K, & \mbox{if } -\frac{3K}{2}\le{\rm Re}\{x_{\rm TD}[m',n']\}< -\frac{K}{2} \\
    ...
  \end{cases}
\end{aligned}
\label{DEMOX12}
\end{equation}
Based on (\ref{DEMOX12}), it is evident that the pdf of ${\rm Re}\{x_{\rm TD}[m',n']\}$ is necessary to characterize ${\rm Re}\left\{{I^{\rm High\ SNR}_{\rm TD}[m',n']}/{|h_1|}\right\}$.\par

Note that $x_{\rm TD}[m',n']$ is obtained in (\ref{Equ2}) by performing IDFT operation on $X_{\rm DD} [m',k]$. If $N$ is significantly large, $x_{\rm TD}[m',n']$ can be regarded as a complex Gaussian distributed variable based on the central limit theorem (CLT), i.e, $x_{\rm TD}[m',n'] \sim {\cal {CN}}(0,{{\cal E}_{X_{\rm DD}}})$ \cite{bxxx}. Based on this, and assuming that $x_{\rm TD}[m',n']$s are independent and identically distributed (i.i.d.) variables, we can deduce that both $\sqrt{\frac{2}{{{\cal E}_{X_{\rm DD}}}}}{\rm Re}\{x_{\rm TD}[m',n']\}$ and $\sqrt{\frac{2}{{{\cal E}_{X_{\rm DD}}}}}{\rm Im}\{x_{\rm TD}[m',n']\}$ are i.i.d. standard normally distributed variables, i.e., $\sqrt{\frac{2}{{{\cal E}_{X_{\rm DD}}}}}{\rm Re}\{x_{\rm TD}[m',n']\}\sim {\cal N}(0,1)$. Based on these and (\ref{DEMOX12}), it can be derived that:
\begin{equation}
\begin{aligned}
&{\rm Pr}\left({\rm Re}\left\{{I^{\rm High\ SNR}_{\rm TD}[m',n']}/{|h_1|}\right\}= \eta K\right)\\
& = {\rm Pr}\left((2\eta-1)\frac{K}{2}\le{\rm Re}\{x_{\rm TD}[m',n']\}<(2\eta + 1)\frac{K}{2}\right)\\
& = \!Q\!\left(\!\sqrt{\frac{2}{{{\cal E}_{X_{\rm DD}}}}}\!(\!2\eta\!-\!1\!)\!\frac{K}{2}\!\right) \!-\! Q\!\left(\!\sqrt{\frac{2}{{{\cal E}_{X_{\rm DD}}}}}\!(\!2\eta \!+\! 1\!)\!\frac{K}{2}\!\right)\!,
\end{aligned}
\label{DEMOX13b}
\end{equation}
where $\eta$ is an integer. We observe from (\ref{DEMOX13b}) that:
\begin{equation}
\begin{aligned}
{\rm Pr}& \left({\rm Re}\left\{{I^{\rm High\ SNR}_{\rm TD}[m',n']}/{|h_1|}\right\}=\eta K\right)\\
& = {\rm Pr}\left({\rm Re}\left\{{I^{\rm High\ SNR}_{\rm TD}[m',n']}/{|h_1|}\right\}= -\eta K\right).
\end{aligned}
\label{NEW39}
\end{equation}
Recall that the distributions of ${\rm Re}\{x_{\rm TD}[m',n']\}$ and ${\rm Im}\{x_{\rm TD}[m',n']\}$ are the same; thus, we can obtain:
\begin{equation}
\begin{aligned}
{\rm Pr}&\left({\rm Re}\left\{{I^{\rm High\ SNR}_{\rm TD}[m',n']}/{|h_1|} = \eta K\right\}\right)\\
& = {\rm Pr}\left({\rm Im}\left\{{I^{\rm High\ SNR}_{\rm TD}[m',n']}/{|h_1|}= \eta K\right\}\right).
\end{aligned}
\label{NEW40}
\end{equation}
Therefore, the mean of $I^{\rm High\ SNR}_{\rm TD}[m',n']/|h_1|$ becomes:
\begin{equation}
\begin{aligned}
&\!{E}\left[I^{\rm High\ SNR}_{\rm TD}\![m',n']/|h_1|\right] \! = 0.
\end{aligned}
\label{DEMOX14a}
\end{equation}
Meanwhile, the variance of $I^{\rm High\ SNR}_{\rm TD}[m',n']/|h_1|$ can be obtained based on (\ref{DEMOX13b})-(\ref{DEMOX14a}) as:
\begin{subequations}
\begin{equation}
\small
\begin{aligned}
&{Var}(I^{\rm High\ SNR}_{\rm TD}[m',n']/|h_1|) = {E}\left[\left|I^{\rm High\ SNR}_{\rm TD}[m',n']/|h_1|\right|^2\right]\\
& = 2{E}\left[{\rm Re}\{I^{\rm High\ SNR}_{\rm TD}[m',n']/|h_1|\}^2\right]\\
& {=}\! \!\sum^{\!\infty\!}_{\eta \!=\! -\infty}\!\!\!2\eta^2K^2\!\left(\!Q\!\left(\sqrt{\frac{2}{{{\cal E}_{X_{\rm DD}}}}}\!(\!2\eta-1\!)\!\frac{K}{2}\!\right)\! \!-\! Q\!\left(\!\sqrt{\frac{2}{{{\cal E}_{X_{\rm DD}}}}}\!(\!2\eta + 1\!)\!\frac{K}{2}\!\right)\!\right)\!
\end{aligned}
\label{DEMOX14b}
\end{equation}
\begin{equation}\small
{=}\! \!\sum^{\!\infty\!}_{\eta \!=\! -\infty}\!\!\!8\eta^2\alpha^2{\cal M}\!\left(\!Q\!\left(\!\sqrt{\!\frac{3{\cal M}}{{\cal M}\!-\!1}}\!(\!2\eta\!-\!1)\!{\alpha}\!\right)\! \!-\! Q\!\left(\!\sqrt{\!\frac{3{\cal M}}{{\cal M}\!-\!1}}\!(\!2\eta + 1)\!{\alpha}\!\right)\!\right)\!,
\label{DEMOX14bb}
\end{equation}
\end{subequations}
where (\ref{DEMOX14b}) and (\ref{DEMOX14bb}) are obtained by substituting (\ref{DEMOX13b}), and (\ref{DEMOX9}) and $K = 2\alpha\sqrt{\cal M}$ into (\ref{DEMOX14b}) and (\ref{DEMOX14bb}), respectively.
Since $I_{\rm DD}[m',k']$ in (\ref{DEMOX6}) is obtained by performing DFT on $I_{\rm TD}[m',n']$, it can also be regarded as a complex Gaussian distributed variable according to the CLT when $N$ is significantly high, i.e., $I_{\rm DD}[m',k']/|h_1|\sim {\cal{CN}}(0,{Var}(I^{\rm High\ SNR}_{\rm TD}[m',n']/|h_1|))$ \cite{bxxx}. Based on the above deductions, (\ref{DEMOX14a}), (\ref{DEMOX14bb}), and according to (3.13) in \cite{b2}, we can obtain the BER associated with the {\it modulo signal loss}, which is given in the following proposition.
\begin{prop}
BER caused by the modulo signal loss for the 4-QAM modulation is given as:
\begin{equation}
{\cal P}^{\rm MSL}_{e,\ {\rm 4-QAM}} {=} Q\left(\sqrt{\frac{2}{{Var}(I^{\rm High\ SNR}_{\rm TD}[m',n']/|h_1|)}}\right).
\label{DEMOX15}
\end{equation}
\end{prop}
According to (\ref{DEMOX15}), we first observe that the BER associated with the {\it modulo signal loss} is independent of the channel fading. Furthermore, we note that the system BER affected by the {\it modulo signal loss} is primarily determined by the scaling factor $\alpha$ in the {\it modified} modulus $K = 2\alpha\sqrt{\cal M}$ and the QAM modulation order $\cal M$. As $\alpha$ increases, ${Var}(I^{\rm High\ SNR}_{\rm TD}[m',n']/|h_1|)$ in (\ref{DEMOX14b}) decreases, resulting in an exponential decrease in ${\cal P}^{\rm{MSL}}_{e}$. Recalling Proposition 1, which indicates that ${\cal P}^{\rm PL}_{e}$ can not be a arbitrary large $\alpha$, we emphasize that the trade-off introduced by the settings of $\alpha$ must be carefully considered when simultaneously minimizing ${\cal P}^{\text{MSL}}_{e}$ and ${\cal P}^{\text{PL}}_{e}$.\par

Finally, we note that since BERs affected by the {\it power loss}, {\it modulo noise loss}, {\it modulo signal loss} are obtained separately for different conditions in Propositions 1, 2, 3, respectively, the system overall BER can not be better than any one of them under different conditions. Considering this, the following theorem for the BER of the proposed design is obtained.\par
\begin{thre}
The BER of the ODDM system with time domain THP for the $4$-QAM modulation is lower bounded by:
\begin{equation}
\begin{aligned}
{\cal P}_{e}\left|_{\alpha,\sigma_w,h_1}\right. & \ge {\cal P}^{\rm LB}_{e},\\
& = {\rm max}\!\left\{\!{\cal P}^{\rm PL}_{e,\ {4}-{\rm QAM}}\!,{\cal P}_{e,\ {4}-{\rm QAM}}^{\rm MNL}\!,{\cal P}^{\rm MSL}_{e,\ {4}-{\rm QAM}}\!\right\}\!,
\end{aligned}
\label{DEMOX23}
\end{equation}
where $h_1 \sim {\cal CN}(0,\sigma^2_{h_1})$, $w \sim {\cal CN}(0,\sigma^2_{w})$, and ${\cal P}^{\rm PL}_{e,\ {4}-{\rm QAM}}$, ${\cal P}_{e,\ {4}-{\rm QAM}}^{\rm MNL}$, and ${\cal P}^{\rm MSL}_{e,\ {4}-{\rm QAM}}$ are derived in Prepositions 1, 2, and 3, respectively.
\end{thre}

\begin{remark}
{\color{black}We note that if the IDFT/DFT operations shown in Fig. \ref{FIG3} are discarded (or replaced by identity transformations), the proposed design will become the single-carrier system with time domain THP \cite{Review1}, i.e., the classical THP system. We clarify that the classical THP system has been well investigated, and it has been shown that BER of such systems is jointly effected by the {\it power loss} and {\it modulo noise loss} \cite{b1}. In addition to the above two losses, interestingly, the BER of the proposed design is additionally affected by the {\it modulo signal loss}, due to the consideration of DD domain data mapping and time domain THP.}
\end{remark}
\begin{remark}
It has been illustrated in \cite{b1} that for classical THP systems, the {\it modulo loss} and {\it power loss} are more significant for small modulation orders, while their impact is negligible for large modulation orders. However, it will be demonstrated in the next section that this conclusion is invalid in our ODDM system with time domain THP. This is due to the {\it modulo signal loss} which was not present in classical THP.\par
\end{remark}

\subsection{Performance Analysis for ODDM with 16-QAM Modulation}
\label{SECTIONIVB}
Based on the Propositions derived in Section \ref{SECTIONIVA} for 4-QAM modulation, in this subsection, we can derive the following corollaries for the ODDM system with time domain THP under for 16-QAM modulation.\par

First, recall that the lower bound for the BER associated with the {\it power loss} for 4-QAM, ${\cal P}^{\rm PL}_{e,\ {\rm 4-QAM}}$, was derived by considering that the modified modulo interference could be ignored when the modified modulus is sufficiently large. Using a similar process and considering the average energy of the 16-QAM symbols is $ {{\cal E}_{X_{\rm DD}}}= \frac{15}{16\alpha^2}{{\cal E}_{x^{\rm THP}_{\rm TD}}}$, we can obtain the BER associated with the {\it power loss} for 16-QAM, ${\cal P}^{\rm PL}_{e,\ {\rm 16-QAM}}$, which is given in the following Corollary. 
\begin{coro}
BERs caused by the power loss for the 16-QAM modulation can be obtained as:
\begin{equation}
\begin{aligned}
& {\cal P}^{\rm PL}_{e,\ {\rm 16-QAM}} \ge \frac{1}{2} - \frac{3}{8}\sqrt {\frac{{{\frac{{\sigma _{{h_1}}^2{\cal E}_{X_{\rm DD}}}}{\sigma_w^2}} }}{{10 + {\frac{{\sigma _{{h_1}}^2{\cal E}_{X_{\rm DD}}}}{\sigma_w^2}} }}}  \\
& - \frac{1}{4}\sqrt {\frac{{9{\frac{{\sigma _{{h_1}}^2{\cal E}_{X_{\rm DD}}}}{\sigma_w^2}} }}{{10 + 9{\frac{{\sigma _{{h_1}}^2{\cal E}_{X_{\rm DD}}}}{\sigma_w^2}} }}} + \frac{1}{8}\sqrt {\frac{{5{\frac{{\sigma _{{h_1}}^2{\cal E}_{X_{\rm DD}}}}{\sigma_w^2}} }}{{2 + 5{\frac{{\sigma _{{h_1}}^2{\cal E}_{X_{\rm DD}}}}{\sigma_w^2}} }}}\\
& = \frac{1}{2} - \frac{3}{8}\sqrt {\frac{{3\sigma _{{h_1}}^2{\rm SNR}}}{{32{\alpha ^2} + 3\sigma _{{h_1}}^2{\rm SNR}}}}  - \frac{1}{4}\sqrt {\frac{{27\sigma _{{h_1}}^2{\rm SNR}}}{{32{\alpha ^2} + 27\sigma _{{h_1}}^2{\rm SNR}}}} \\
 & + \frac{1}{8}\sqrt {\frac{{75\sigma _{{h_1}}^2{\rm SNR}}}{{32{\alpha ^2} + 75\sigma _{{h_1}}^2{\rm SNR}}}},
\end{aligned}
\label{DEMOX1016Q}
\end{equation}
{where} $\alpha$ is sufficiently large, $h_1 \sim {\cal{CN}}(0,\sigma^2_{h_1})$, and ${\rm Re}\{x^{\rm THP}_{\rm TD}[m,n]\},{\rm Im}\{x^{\rm THP}_{\rm TD}[m,n]\} {\sim} {\cal U}(-\alpha\sqrt{\cal M},\alpha\sqrt{\cal M})$.
\end{coro}
Second, recall that Proposition 2 was {derived} based on (\ref{DEMOX16}) and Fig. \ref{FIG4}. Similarly, we next {derive} the BER associated with the {\it modulo noise loss} for 16-QAM, ${\cal P}^{\rm MNL}_{e,\ {\rm 16-QAM}}$. We consider the same case indicated by (\ref{DEMOX16}). Next, we investigate the errors pattern associated with the {\it modulo noise loss} of 16-QAM DD domain symbols in Fig. \ref{FIGA1}, which is given in the Appendix A. Then, by following the steps outlined in Appendix A, BER due to the {\it modulo noise loss} under the 16-QAM modulation can be obtained, which is given next in the Corollary.
\begin{coro}
Under $h_1\sim {\cal CN}(0,\sigma^2_{h_1})$ and $w[m,k]\sim {\cal CN}(0,\sigma^2_w)$, a lower bound for the BER due to the {\it modulo noise loss} under the 16-QAM modulation is given by:
\begin{equation}
\begin{aligned}
& {\cal P}_{e,\ {\rm 16-QAM}}^{\rm MNL} \ge 2\left({\cal P}_{\rm case-a}{+}{\cal P}_{\rm case-b}{+}{\cal P}_{\rm case-c}{+}{\cal P}_{\rm case-d}\right).
\end{aligned}
\label{BER16QAM}
\end{equation}
where ${\cal P}_{\rm case-a}$, ${\cal P}_{\rm case-b}$, ${\cal P}_{\rm case-b}$, and ${\cal P}_{\rm case-d}$ are given by (\ref{Equ53}), (\ref{Equ54}), (\ref{Equ55}), and (\ref{Equ56}) in Appendix A, respectively. 
\end{coro}
Third, recall that the BER associated with the {\it modulo signal loss} for 4-QAM, ${\cal P}^{\rm MSL}_{e,\ {\rm 4-QAM}}$, was derived based on the assumption that the noise term $\tilde{w}[\bar{m}]$ of (\ref{DEMOX4}) is negligible. Using a similar process, while considering $I_{\rm DD}[m',k']/|h_1|\sim {\cal{CN}}(0,{Var}(I^{\rm High\ SNR}_{\rm TD}[m',n']/|h_1|))$ and (8.15) in \cite{bMS}, we derive the BER associated with the {\it modulo signal loss} for 16-QAM, ${\cal P}^{\rm MSL}_{e,\ {\rm 16-QAM}}$, which is given in the following Corollary. 
\begin{coro}
When noise is negligible, BER caused by the modulo signal loss for the 16-QAM modulation is:
\begin{equation}
{\cal P}^{\rm MSL}_{e,\ {\rm 16-QAM}} {=} \frac{3}{4}Q\left(\sqrt{\frac{{{\cal E}_{X_{\rm DD}}}}{{{{5}{Var}(I^{\rm High\ SNR}_{\rm TD}[m',n']/|h_1|)}}}}\right).
\label{DEMOX15B}
\end{equation}
\end{coro}
Finally, we note that similar to that in Theorem 1, the lower bounds for the BER for 16-QAM can be obtained based on Corollaries 1, 2, and 3. Furthermore, we specify that by following steps similar to the above, the analytical results for the proposed design under any other modulation, e.g., 64-QAM, BPSK, etc., can be obtained.


\section{Numerical Results}
In this section, we present numerical results to evaluate the performance of the proposed ODDM system with time domain THP. For the SRN pulse in the ODDM system, we consider the square-root-raise-cosine (SRRC) pulse with the roll-off factor $\beta = 0.1$ and ${\cal Q}=20$. {\color{black}Unless otherwise specified, other} ODDM system parameters are $\Delta f=15$ kHz, $M=512$, $N=64$, and the carrier frequency $f_c=6$ GHz. Moreover, two typical high mobility models are considered, namely the EVA model \cite{bb13} and the high-speed railway (HSR) model \cite{bb14}. In the EVA model, gains of the nine propagation paths follow Rayleigh fading, as defined in \cite{bb13}, and the maximum Doppler shift is 1000 Hz. In the HSR model, gains of the four propagation paths follow Rician fading with $K$-factor defined as in \cite{bb14}, and the maximum Doppler shift is 2000 Hz. 

\subsection{BER under Different Scaling Factor $\alpha$}
We first investigate the influence of the scaling factor $\alpha$ on the BER of our ODDM system with time domain THP. To this end, in Fig. \ref{FIGJ1}, the BER performance in the EVA channel with 4-QAM modulation and SNR $= 30$ dB is plotted. Therein, first, {\color{black}the simulated BERs under different $N$ and $M = 512$} are plotted. Second, the BER lower bound derived in Theorem 1 is plotted. The decomposed components of the BER lower bound, derived in Propositions 1, 2, and 3, are also plotted.\par
\begin{figure}[t]
  \centering
  \includegraphics[height = 160pt, width=.4\textwidth]{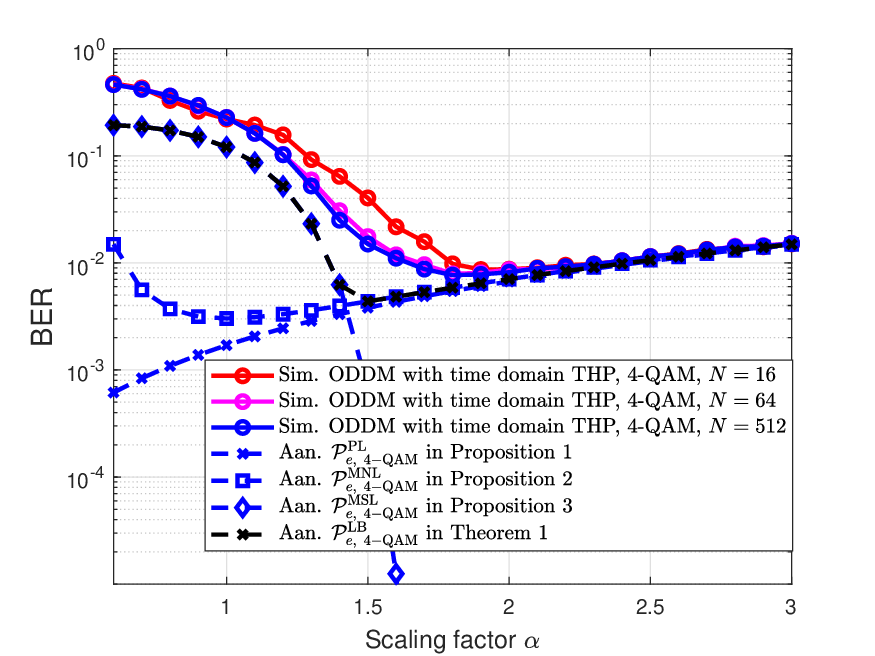}
  \caption{{\color{black}BER of the ODDM system with time domain THP under 4-QAM modulation versus the scaling factor $\alpha$ in the modified modulus $K$.}}\label{FIGJ1}
  \vspace{-1em}
\end{figure}
First, we observe that the analytical result in Theorem 1 acts as a close lower bound for the simulation results of the ODDM system with time domain THP, especially for large scaling factors. This shows the correctness of our BER analysis in Section IV. Second, we focus on the low $\alpha$ regions, i.e., $0.6\le\alpha\le1.8$, and observe that the {\it modulo signal loss} ${\cal P}^{\rm MSL}_{e}$ derived in Proposition 3 dominates the system BER performance. This is because the possibility of ${\rm Re}\{x^{\rm}_{\rm TD}[m,n]\} >2\alpha\sqrt{\cal M}$ and ${\rm Im}\{x^{\rm}_{\rm TD}[m,n] > 2\alpha\sqrt{\cal M}\}$ is high when $\alpha$ is small. {\color{black}Meanwhile, we also note that the derived BER lower bound ${\cal P}^{\rm LB}_{e}$ in this low $\alpha$ region is not as tight as that in the high $\alpha$ region. In detail, it can be observed that when $N = 64$, the simulation results are closer to the theoretical bound than the case with $N = 16$. This is because when $N = 16$, the assumption that the time-domain symbols before THP follow a Gaussian distribution does not hold since $N$ is too small to make the central limit theorem valid. Also, it can be observed that the simulation results under $M = 512, N = 64$ are very close to those under $M = 512, N = 512$. Therefore, it can be inferred that the Gaussian distribution assumption used to derive Proposition 3 can be regarded as valid when $M = 512, N = 64$. Based on these, we can deduce that the gap between the simulation and theoretical curves in the low $\alpha$ region results from that the three types of losses affect the system BER at the same time, while Theorem 1 only considers the most serious one. In other words, Theorem 1 gives a lower bound, but it is not necessarily a tight bound.} {Third, we observe that the lowest BER for the proposed design is achieved in the medium $\alpha$ region, i.e., $1.8< \alpha \le 2.2$, where the BER is jointly effected by all the three kinds of losses.} Finally, we observe that ${\cal P}_{e}^{\rm MNL}$ is lower bounded by ${\cal P}^{\rm PL}_{e}$, which has been predicted by (\ref{DEMOX37}) and shows that ${\cal P}_{e}^{\rm MNL}\ge{\cal P}^{\rm PL}_{e}$ holds.\par

\begin{figure}[t]
  \centering
  \includegraphics[height = 160pt, width=.4\textwidth]{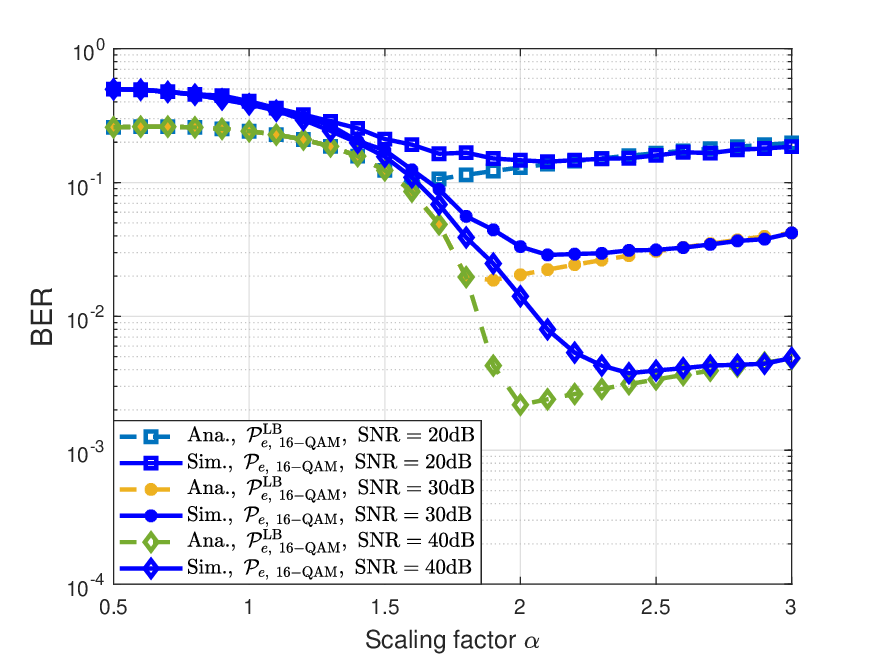}
  \caption{\color{black}BER of the ODDM system with time domain THP under 16-QAM modulation versus the scaling factor $\alpha$ in the EVA channel.}\label{FIGJ3}
  \vspace{-1em}
\end{figure}
In Fig. \ref{FIGJ3}, we examine the BER of the proposed design considering 16-QAM modulation, under varying $\alpha$ and SNR. To this end, we plot the simulated BERs ${\cal P}_{e,\ \rm 16-QAM}$ and the derived BER lower bound ${\cal P}^{\rm LB}_{e,\ \rm 16-QAM}$ in Theorem 1 based on Corollaries 1, 2, and 3.\par
Similar to the observations in Fig. \ref{FIGJ1}, we find that (i) the analytical results for our system with 16-QAM modulation provide close lower bounds for its simulation results, and (ii) the system BER is lower bounded by the {\it modulo signal loss} in the low $\alpha$ regions for different SNRs, i.e., $\alpha < 1.5$, and (iii) the system BER is lower bounded by the {\it modulo noise loss} in the high $\alpha $ regions. Second, we observe that the value of $\alpha$ that minimizes ${\cal P}_{e,\ \rm 16-QAM}$ is different for different SNRs. For example, $\alpha = 2.1$ could minimize ${\cal P}_{e,\ \rm 16-QAM}$ when SNR = 20 dB. Differently, {$\alpha$} that minimizes ${\cal P}_{e,\ \rm 16-QAM}$ under SNR = 40 dB is  $\alpha= 2.4$. This shows that while considering the trade-off between the impacts of modulo signal loss and modulo noise loss, $\alpha$ needs to be carefully set to achieve the best system BER performance under different channel environments, QAM modulation orders, and SNRs.\par

\begin{figure}[t]
  \centering
  \includegraphics[height = 160pt, width=.4\textwidth]{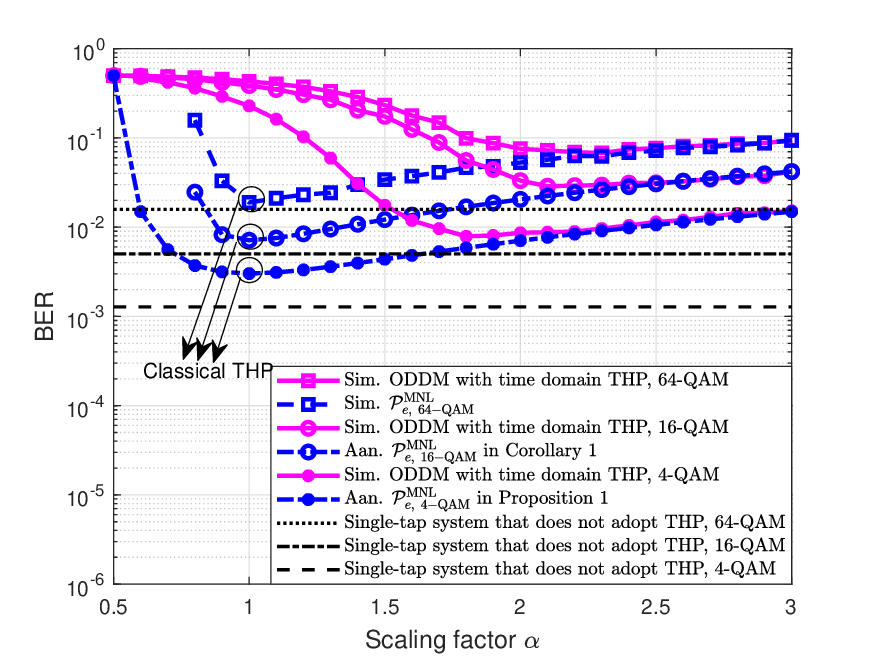}
  \caption{\color{black}BER of different systems versus the scaling factor $\alpha$ for different QAM modulation orders.}\label{FIGJ2}
  \vspace{-1em}
\end{figure}
In Fig. \ref{FIGJ2}, we illustrate the impacts of QAM modulation order on the ODDM system with time-domain THP under varying $\alpha$ and SNR = 30 dB. To this end, we present the simulated BERs for our ODDM system with time domain THP under the EVA channel with 4-QAM, 16-QAM, and 64-QAM, respectively. For comparison, we first present the simulated BERs of the single-tap system that does not adopt THP. Second, we plot the BERs associated with the classical THP system, where both data mapping and THP are performed in the time domain.\par

First, we note that as the modulation order increases, the performance loss brought by the THP becomes negligible for the classical THP systems compared to the single-tap system that does not adopt THP. This is because the higher order modulation reduces the proportion of symbols that are likely to experience the {\it modulo loss} in the high SNR region \cite{b1}. {\color{black} On the contrary, as the modulation order increases, the performance loss due to THP is always significant for our ODDM system with time domain THP design, as a result of considering DD domain data mapping for utilizing the stationarity of CSF and time domain THP for low-complexity ISI pre-cancellation.} This is due to the fact that the {\it modulo signal loss} can only be neglected in the large $\alpha$ region, e.g., $\alpha > 2$ for the system with 16-QAM and $\alpha > 1.5$ for the system with 4-QAM. In this high $\alpha$ region, the {\it power loss} has already introduced significant performance loss. {\color{black} As a result, the BER of our design could be higher than those of both classical THP systems and the single-tap system without THP.} Based on the above observations, we can deduce that the BER performance of the proposed design can be significantly improved if the {\it modulo signal loss} could be mitigated in the small $\alpha$ regions. As potential strategies, the time domain symbols for which the THP will be performed, can be clipped. This consideration is similar to that considered in several OFDM systems \cite{CLIP}, and will be investigated in our further research efforts.\par

\subsection{BER under Different SNRs}
\begin{figure}[t]
  \centering
  \includegraphics[height = 160pt, width=.4\textwidth]{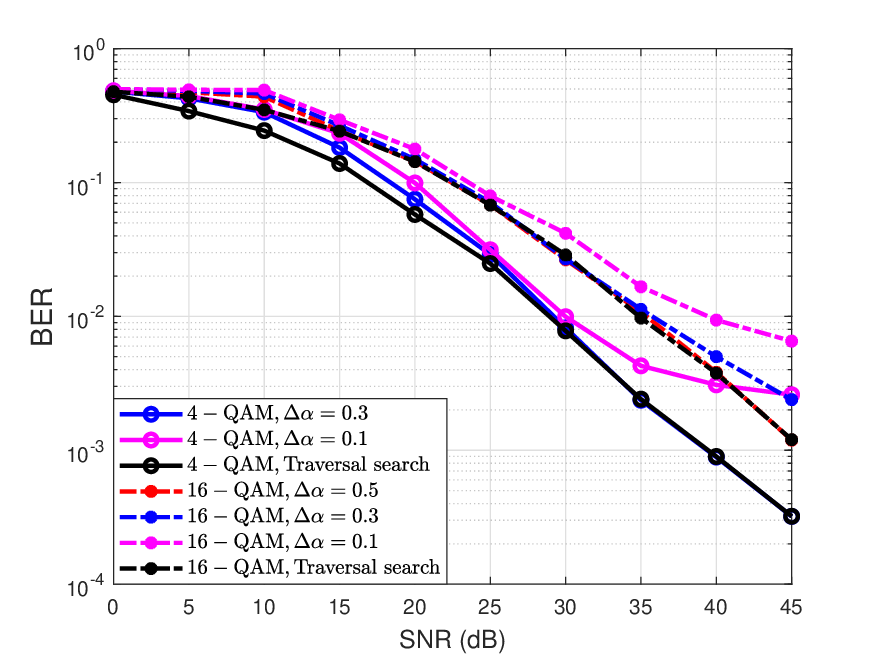}
  \caption{\color{black}{Minimum BER of the proposed ODDM system with time domain THP with different searching intervals for $K^{\rm sim}_{\rm opt}$ under different SNRs and the HSR channel.}}\label{FIGNA}
    \vspace{-1em}
\end{figure}
	{\color{black}According to Figs. \ref{FIGJ1} - \ref{FIGJ2}, it can be observed that the modified modulus $K$ is of great importance to achieve good BER performance for the proposed design. Denote the $K$ which minimizes the analytical and simulated BERs as $K^{\rm ana}_{\rm opt}$ and $K^{\rm sim}_{\rm opt}$, respectively, e.g., $K^{\rm ana}_{\rm opt} = 3\sqrt{\cal M}$ and $K^{\rm ana}_{\rm opt} = 3.6\sqrt{\cal M}$ in Fig. \ref{FIGJ1}, which could be different. Meanwhile, it will be complex to find $K^{\rm sim}_{\rm opt}$ through simulation under different SNRs with $K$ located in an arbitrary interval, e.g., $K\in [1.2\sqrt{\cal M},6\sqrt{\cal M}]$ in Figs. \ref{FIGJ1} - \ref{FIGJ2}. However, interestingly, we can observe that $K^{\rm sim}_{\rm opt}$ locates in the neighborhood of $K^{\rm ana}_{\rm opt}$. Thus, we can define $\Delta \alpha$ and find the minimum BER of our proposed system through simulation, under the searching regions for $K^{\rm sim}_{\rm opt}$ with $K^{\rm sim}_{\rm opt} \in [K^{\rm ana}_{\rm opt},K^{\rm ana}_{\rm opt}+\Delta \alpha]$.\par}

	{\color{black}The obtained minimum BERs of the proposed design with different searching intervals for $K^{\rm sim}_{\rm opt}$ are given in Fig. \ref{FIGNA}. In Fig. \ref{FIGNA}, ``Traversal search'' refers to the case that $K^{\rm sim}_{\rm opt}$ is traverse searched within a longer interval, e.g., $K\in [1.2\sqrt{\cal M},6\sqrt{\cal M}]$ in Figs. \ref{FIGJ1} - \ref{FIGJ2}, whose BER can be regarded as the minimum BER that the proposed design could attain.\par}

{\color{black}First, it can be observed that when $\alpha$ is small, e.g., $\Delta \alpha = 0.1$, the minimum BER of the proposed design will encounter an error floor at high SNR regions. This is because the theoretical BER lower bound ${\cal P}^{\rm LB}_{e}$ is not tight. Second, we can find that for the proposed design under the EVA channel, whether 4-QAM or 16-QAM modulation is used, $\alpha=0.5$ is sufficient to obtain the minimum BER of the system at high SNRs. Therefore, the $K^{\rm ana}_{\rm opt}$ obtained according to the analytical results could provide a reference to set the modified modulus $K$ efficiently.}

\begin{figure}[t]
  \centering
  \includegraphics[height = 160pt, width=.4\textwidth]{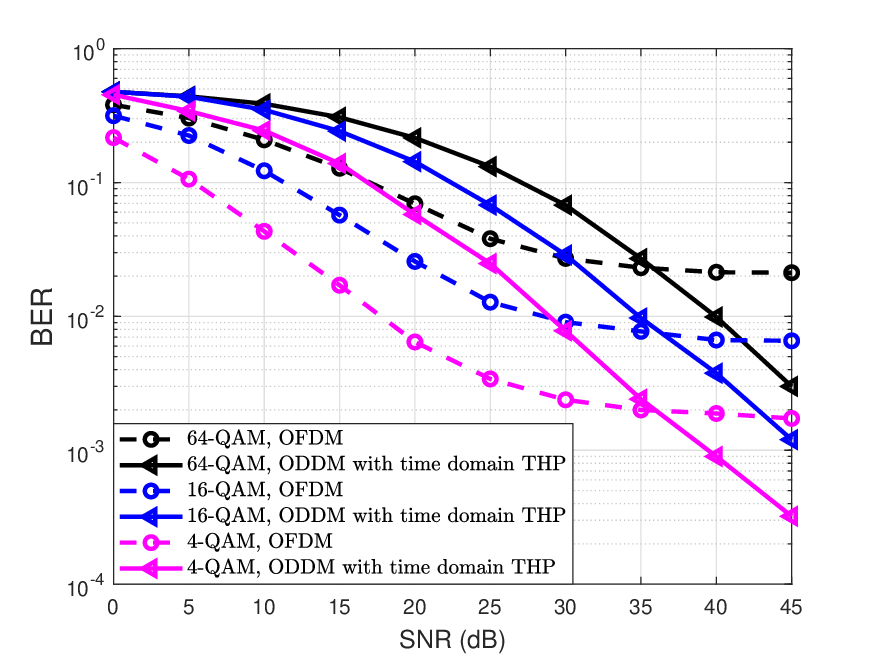}
  \caption{\color{black}{BER comparison of the proposed ODDM system with time domain THP with the OFDM system with single-tap equalizer.}}\label{FIGNN5C}
    \vspace{-1em}
\end{figure}
{\color{black}Next, in Fig. \ref{FIGNN5C}, we evaluate the BER of the ODDM system with time domain THP by comparing it to with the OFDM system with the frequency domain single-tap equalizer, for the fair comparison w.r.t. the equalizer complexity. Therein, the same system parameters and the HSR channel environment are considered for two systems. The minimum BERs of the proposed ODDM system with time domain THP under different SNRs are simulated {\color{black}according to Fig. \ref{FIGNA}} and plotted.}\par

{We can observe from Fig. \ref{FIGNN5C} that the BER of the proposed scheme decreases monotonically as SNR increases for all QAM modulation orders and channel environments. Differently, the BERs of the OFDM system experience error floors in the high SNR regions. This is because for the ODDM system with time domain THP, on one hand, the ISI is pre-cancelled by THP at the transmitter; on the other hand, impacts of the {\it modulo signal loss}, {\it modulo power loss}, and {\it modulo noise loss} introduced by THP weaken as the SNR increases. These leads to the BER of the proposed scheme to decrease monotonically as SNR increases. However, for the OFDM system with single-tap equalizer, the ICI can not be addressed which determines the lowest signal-to-interference-and-noise ratio of the system.\par

Subsequently, we compare the BER of the ODDM system with time domain THP (same as those in Fig. \ref{FIGNN5C}) to {\color{black}benchmarks based on ODDM} in Fig. \ref{FIGN5}. In particular, as benchmarks, we consider (i) {\color{black}the ODDM system with the MP detector \cite{bb4}} and (ii) the ODDM system with the MRC equalizer \cite{bb15}. We note that in the ODDM system with MRC, the ISI is equalized at the receiver through the serial interference cancellation (SIC) \cite{bb15}, which is similar to the proposed scheme where the ISI is pre-cancelled at the transmitter successively by THP. {\color{black}Besides, the ODDM system with the MP detector considers the sparsity of the DD equivalent channel matrix shown in (\ref{Equ13}), models the ISI as a Gaussian variable, and performs iterative detection through probability propagation between observation nodes and variable nodes \cite{bb4}.}

\begin{figure}[t]
  \centering
  \includegraphics[height = 160pt, width=.4\textwidth]{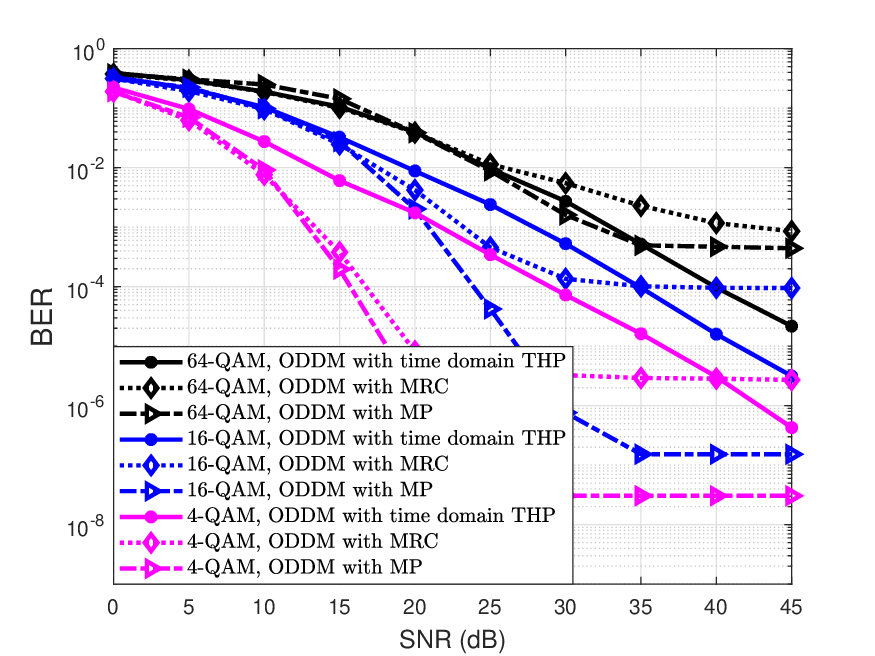}
  \caption{\color{black}{BER comparison of the proposed ODDM system with time domain THP with those of the ODDM system with MP detector and the ODDM system with MRC under the HSR channel.}}\label{FIGN5}
    \vspace{-1em}
\end{figure}
{\color{black}Compared to the proposed design, we first observe from Fig. \ref{FIGN5} that the BERs of the ODDM systems with MP and MRC experience error floors in the high SNR regions.} {\color{black} Different from the reason that the OFDM system meets the error floor in Fig. \ref{FIGNN5C}, the error floor of the ODDM system with MP detector results from that the Gaussian approximation assumption for ISI may not be satisfied; and the error floor of} the ODDM with MRC results from the imperfect iteration input, iteration quantization error, and error propagation.\par
Second, we observe in Fig. \ref{FIGN5} that, in the low SNR regions and 4-QAM modulation, {\color{black}the ODDM systems with MRC and MP} outperform the ODDM system with time domain THP in terms of BER. This is because {\color{black}ODDM systems with MRC and MP} could utilize multipath diversity. On the contrary, ODDM with time domain THP cannot utilize multipath diversity since the ISI is pre-cancelled. Nevertheless, as the increase of SNR, e.g., SNR$>35$ dB under 4-QAM modulation, the BER of the proposed scheme is lower than {\color{black}the ODDM system with MRC}. {\color{black}Moreover, the BER of the ODDM system with THP could be better than others for 64-QAM symbols. This is because for ODDM systems with MRC and MP, a larger constellation size requires more decisions to be performed in each iteration, so the impact of error propagation will be more significant. These observations obtained according to Figs. \ref{FIGNN5C} and \ref{FIGN5} show that when the constellation size is large or in the high SNR regions, the BER superiority of the proposed design over benchmarks is better revealed.\par}
\subsection{{\color{black}BER under Imperfect CSF Assumptions}}
\begin{figure}[t]
  \centering
  \includegraphics[height = 160pt, width=.4\textwidth]{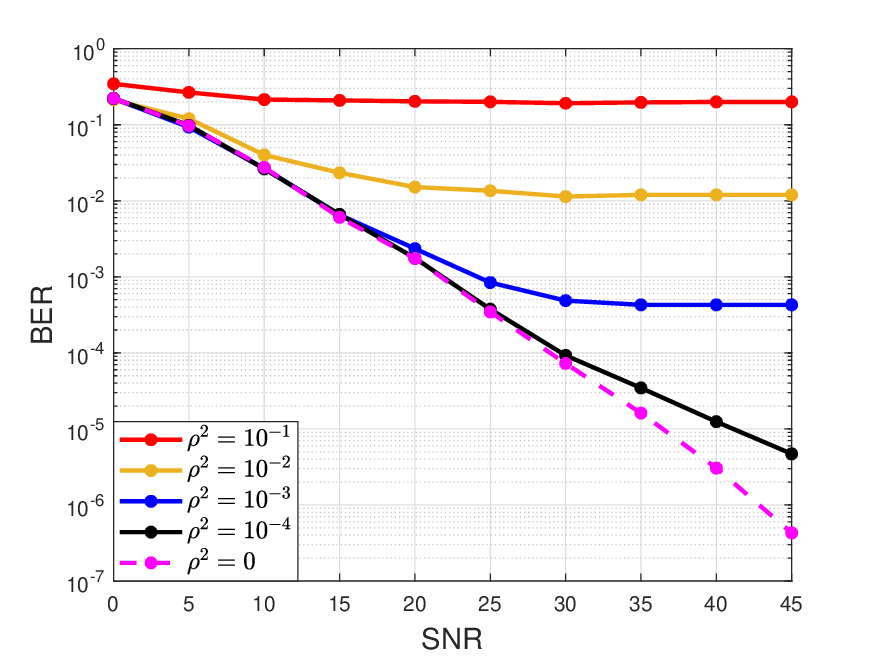}
  \caption{\color{black}{BER of the proposed ODDM system with THP under imperfect CSF assumptions and the HSR channel.}}\label{IMCSF}
  \vspace{-1em}
\end{figure}

{\color{black}Afterwards, Fig. 12 shows the BER of the ODDM system with time domain THP under different degrees of CSF imperfection and the HSR channel environment, where 4-QAM modulation is adopted. The imperfect CSF ${\hat h}(\tau,\nu)$ used for THP is characterized according to \cite{bb4} as:
\begin{equation}
\label{Equ_IMCSF}
  {\hat h}(\tau,\nu) = {h}(\tau,\nu) + w(\tau,\nu),
\end{equation}
where $h(\tau,\nu)$ refers to the perfect CSF and $w(\tau,\nu)$ models the CSF imperfection. Therein, $w(\tau,\nu)\sim {\mathcal CN}(0,\rho^2)$ and $\rho^2$ characterizes the degree of the CSF imperfection{\footnote{\color{black}We note that the Doppler shifts of multipath affect the proposed design by phase rotations in (\ref{THPEqu19B}) and hence, the imperfect Doppler shift can be regarded as part of the imperfect fading coefficient. Thus, we clarify that the model in (\ref{Equ_IMCSF}) accounts for the influences of imperfect estimations in both the channel coefficient and Doppler shifts. Considering the influences of delay errors is beyond the scope of this work, but will be considered in our future works.}}}.\par 

{\color{black}First, we can intuitively observe that as the channel imperfection increases, the BER encountered by the proposed design will become more severe. This is because, as shown by (\ref{THPEqu19B}), the proposed design considers successive interference pre-cancellation. Due to this, channel imperfection makes THP suffer from error-propagation at the transmitter and thus, results in error floors at the receiver. Second, we note that the mean square error (MSE) of ${\hat h}_p$ in (\ref{Equ_IMCSF}) is approximately $MSE = \rho$. For typical DDMC channel
estimators, $MSE \le 10^{-2}$ can be achieved \cite{cs4,cs3}. This shows that even under imperfect CSF, our design can achieve reasonable BERs, e.g., BER of $10^{-5}$ at ${\rm SNR} = 40$ dB. This further validates the significance of our analysis. We clarify that it might of interest to analytically investigate the BER lower bounds under imperfect CSF conditions. This is beyond the scope of this work, but will be considered in our future works.}

\subsection{Complexity Analysis}
\begin{table}[t]
\setlength{\abovecaptionskip}{0.cm}
\setlength{\belowcaptionskip}{-2.cm}
{\bf\caption{Equalization (and Precoding) Complexity of Different ODDM Systems.}}
\begin{threeparttable}
\centering
\begin{tabular}{p{165pt}<{\centering}|p{60pt}<{\centering}}
\hline
{\bf Schemes}&
\bf{Complexity}\\
\hline
Proposed ODDM system with time domain THP&
${\cal O}(NMP)$\\
\hline
ODDM system with MRC \cite{bb15}&
${\cal O}(n_{\rm iter}NM(P+log_2{N}))$\\
\hline
ODDM system with MP \cite{bb4}&
${\cal O}(n_{\rm iter}NMP{\cal M})$\\
\hline
ODDM system with MMSE equalizer \cite{Kehan}&
${\cal O}(N^3M^3)$\\
\hline
\end{tabular}
\label{tab2}
\begin{tablenotes}
 \footnotesize
  \item[*] $n_{\rm iter}$ refers to the number of iterations.
 \end{tablenotes}
  \end{threeparttable}
\end{table}
Finally, in Table \ref{tab2}, we compare the equalization (and precoding) complexity of the ODDM system with THP with several benchmarks. {\color{black}For the proposed design, the complexity of THP depends on (\ref{THPEqu19B}) and (\ref{THPEqu19A}), whose complexity are ${\cal O}(PMN)$ and ${\cal O}(MN)$, respectively. Besides, the complexity of the equalizer in (\ref{DEMOST}) is ${\cal O}(MN)$. Based on these, the overall equalization and precoding complexity of the proposed ODDM system with time domain THP can be found as ${\cal O}(PMN)$.\par}

First, we observe that the proposed ODDM system with time domain THP shows a significant reduction in the equalization complexity compared to the classical ODDM equalizers, i.e., MMSE, MP, and MRC. Besides, the complexity of the proposed scheme is not effected by the QAM modulation order. {\color{black}Second, we observe that the equalization complexity of ODDM system with MRC or MP may be comparable to that of our proposed ODDM system with time domain THP. Despite this, we note that the ODDM system with MRC or MP experiences an error floor, especially for the case with high modulation order. We specify that although the error floor could be mitigated by adjusting the damping factor, input, and decision rules during the iteration, the equalizer complexity would be further sacrificed.}\par
\section{Conclusion}
This article proposed to adopt time domain THP for the ODDM system to make the DD domain single-tap equalizer feasible, thereby significantly reducing the equalization complexity. First, we proposed the ISI pre-cancellation for the ODDM system over the LTV channel. Second, a {\it modified modulo operation} was proposed to realize the joint DD domain data modulation and time domain ISI pre-cancellation. We then analyzed the performance losses resulting from time domain THP, where the BER associated with the {\it power loss}, {\it modulo noise loss}, and {\it modulo signal loss} were discussed. Based on the analytical results, BER lower bounds of our proposed system were obtained for 4-QAM and 16-QAM modulations. In the end, through numerical simulations, we verified the theoretical analysis and showed the BER superiority of the proposed design over OFDM systems with single-tap equalizer and ODDM systems with MRC equalizer in high SNR and channel-diversity-limited scenarios.\par

As future works, several key aspects related to our design will be investigated in detail. Firstly, the channel diversity has not been explored yet by our ODDM system with time domain THP. This investigation can be crucial to improve BER performance, especially in the low SNR region. Secondly, this article assumes that the channel spreading function is perfectly known by the transmitter, while strategies for the imperfect case should be studied. Thirdly, the {\it modulo signal loss} needs to be mitigated, for which the time domain signal distortion approaches can be investigated.

\appendices
\section{BER Associated with the Modulo Noise Loss for 16-QAM Modulation}

We note that the real and imaginary components of the 16-QAM symbols can be regarded as 4-PAM symbols. Since the distributions of ${\rm Re}\{X_{\rm DD} [m',k']{+}\frac{{w}[m',k']}{|h_1|}\}$ and ${\rm Im}\{X_{\rm DD} [m',k']{+}\frac{{w}[m',k']}{|h_1|}\}$ in (\ref{DEMOX16}) are same, we next focus on ${\rm Re}\{X_{\rm DD} [m',k']{+}\frac{{w}[m',k']}{|h_1|}\}$ similar to Fig. \ref{FIG4}. To calculate BER, two bits of 4-PAM symbols need to be considered. This necessitates the investigation of four cases as demonstrated in Fig. \ref{FIGA1} based on the symmetry of bits $0$ and $1$. Therein, the regions of ${\rm Re}\{X_{\rm DD} [m',k']{+}\frac{{w}[m',k']}{|h_1|}\}$ which result in the {\it modulo noise loss} are colored in red. As shown by Fig. \ref{FIGA1}(a), if ${\rm Re}\{X_{\rm DD}[m,k]\} = -3$ is transmitted, the BER associated with the first bit $1$ is calculated as:
\begin{equation}
\small
{\cal P}_{\rm case-a} = \frac{1}{8}E_{h_1}[\sum_{\beta}{\cal F}({( {\beta  - 1})K+3}\!,\!({\beta-\frac{1}{2}}) K+3)],
\label{Equ53}
\end{equation}
where $\beta$ is an integer, and ${\cal F}\left(a,b\right)$ is same as that defined in (\ref{Equ42}). Similarly, if ${\rm Re}\{X_{\rm DD}[m,k]\} = -1$ is transmitted, the BER associated with the first bit $1$ is calculated according to Fig. \ref{FIGA1}(b) as:
\begin{equation}
\small
\!{\cal P}_{\rm case-b}\! =\! \frac{1}{8}E_{h_1}[\sum_{\beta}{\cal F}({( {\beta  - 1} )K+1},({\beta-\frac{1}{2}}) K+1)].\!
\label{Equ54}
\end{equation}
Moreover, if ${\rm Re}\{X_{\rm DD}[m,k]\} = -3$ is transmitted, the BER associated with the second bit $1$ is calculated according to Fig. \ref{FIGA1}(c) as:
\begin{equation}
\small
\!{\cal P}_{\rm case-c} \!=\! \frac{1}{8}E_{h_1}[\sum_{\beta}{\cal F}({( {\beta  - 1} )K+1},({\beta-1}) K+5)].\!
\label{Equ55}
\end{equation}
Additionally, if ${\rm Re}\{X_{\rm DD}[m,k]\} = -1$ is transmitted, the BER associated with the second bit $0$ is calculated according to Fig. \ref{FIGA1}(d) as:
\begin{equation}
\small
{\cal P}_{\rm case-d} = \frac{1}{8}E_{h_1}[\sum_{\beta}{\cal F}({( {\beta  - 1} )K+3},{\beta} K-1)].
\label{Equ56}
\end{equation}
Finally, based on (\ref{Equ53})-(\ref{Equ56}) and the symmetry of bits and real and imaginary components of 16-QAM symbols, a lower bound of ${\cal P}_{e,\ {\rm 16-QAM}}^{\rm MNL}$ is obtained as in (\ref{BER16QAM}).
\begin{figure}[t]
  \centering
  {\includegraphics[width=.45\textwidth]{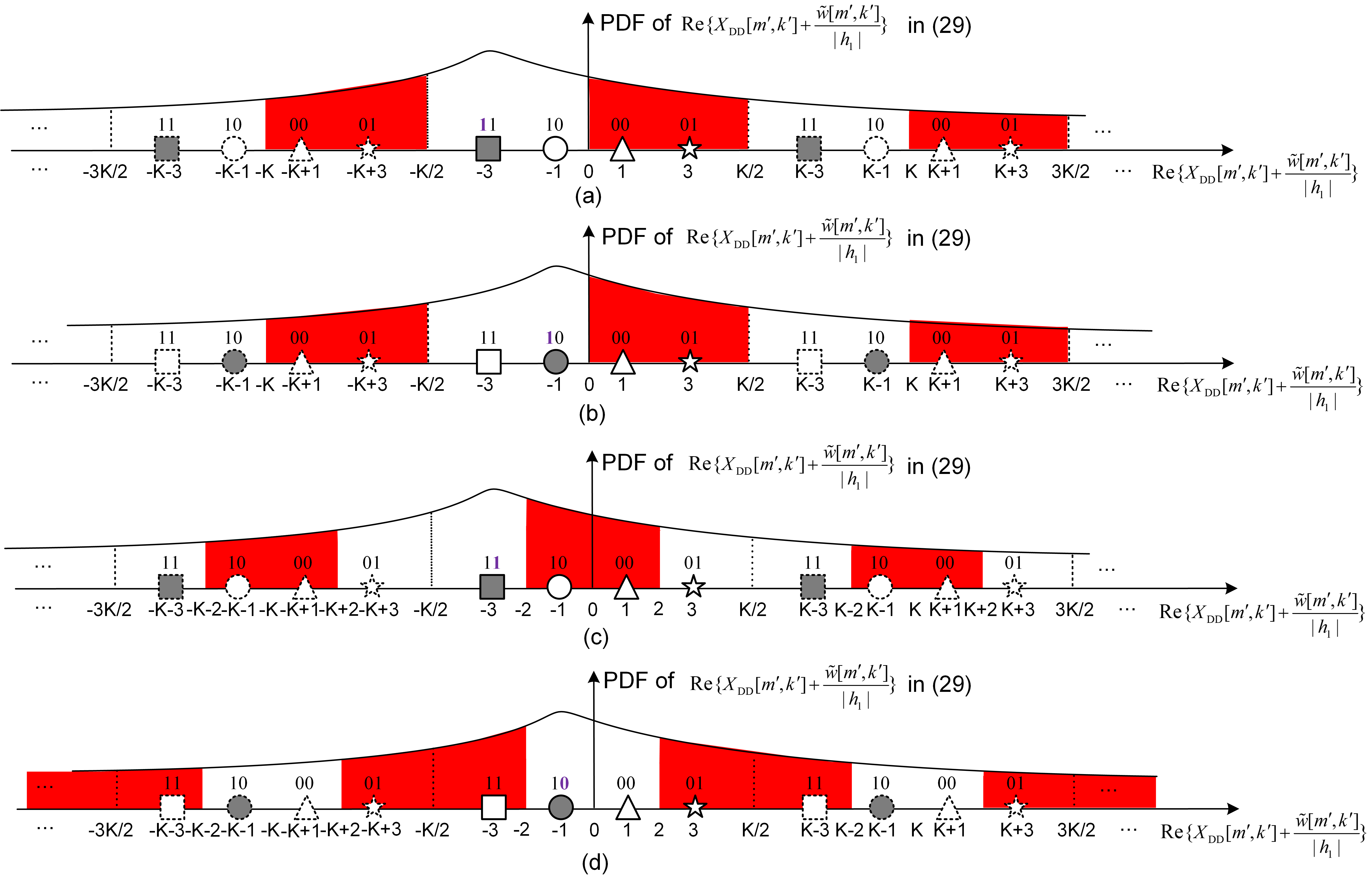}}
  \caption{Illustration of ${\rm Re}\{X_{\rm DD} [m',k']{+}\frac{{w}[m',k']}{|h_1|}\}$ values in (\ref{DEMOX16}) that lead to {\it modulo noise loss} with $|h_1| {=} 1$. Here, the ${\rm Re}\{{X_{\rm DD} [m',k']} \}$ is $-3, -1, -3, -1$ in (a) - (d), respectively. Values of ${\rm Re}\{X_{\rm DD} [m',k']{+}\frac{{w}[m',k']}{|h_1|}\}$ that result in errors are shadded in red.}\label{FIGA1}
  \vspace{-1.5em}
\end{figure}
{}

\begin{IEEEbiography}[{\includegraphics[width=1in,height=1.25in,clip,keepaspectratio]{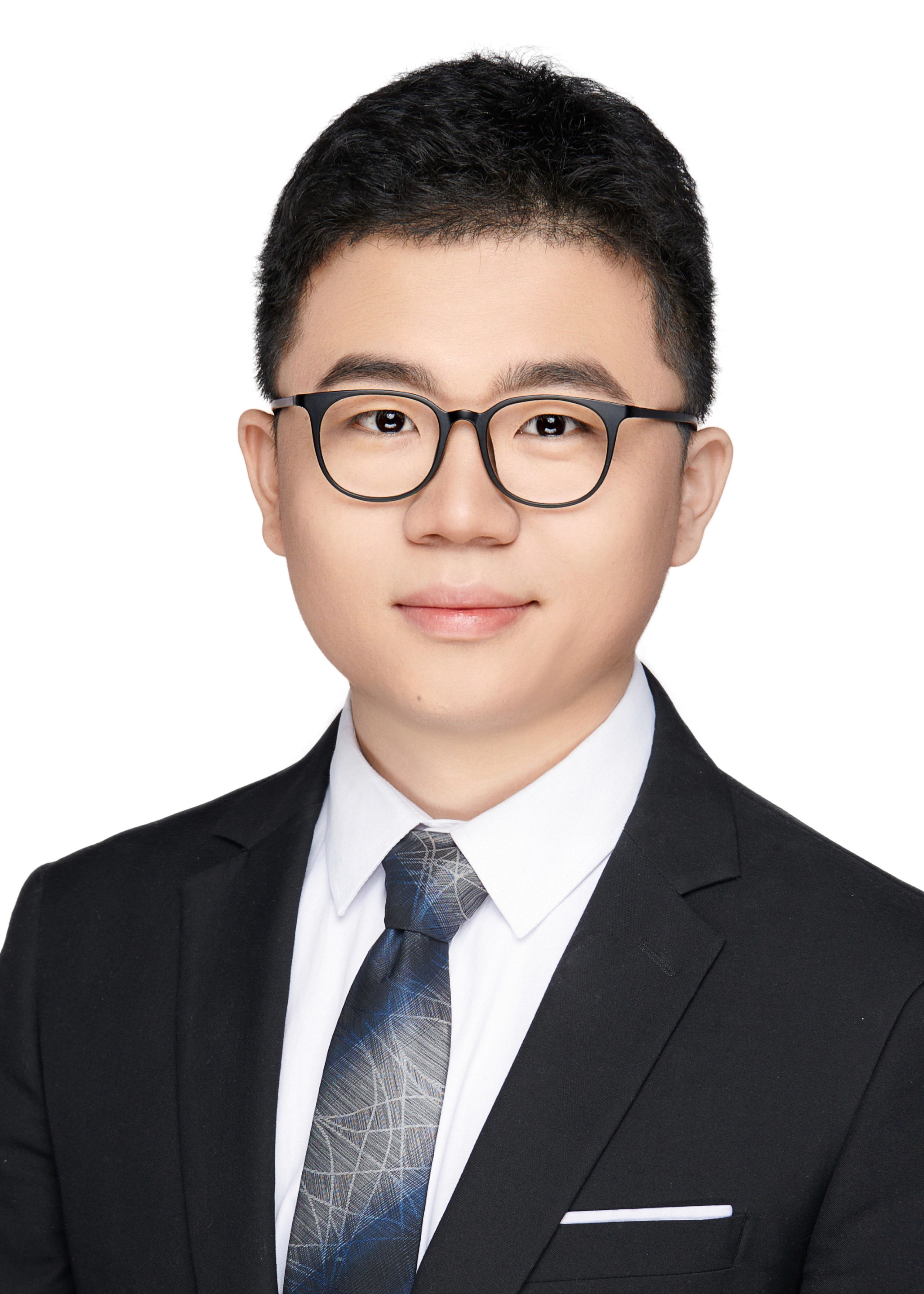}}]{Yiyan Ma}
(Member, IEEE) received the B.S. degree in applied physics and Ph.D. degree in information and communication engineering from Beijing Jiaotong University, Beijing, China, in 2019 and 2024, respectively. Currently, he is a lecturer and Postdoctoral fellow at the School of Electronic and Information Engineering, Beijing Jiaotong University, Beijing, China. His current research interests include the delay-Doppler (DD) domain multi-carrier modulation, DD domain channel characteristics, and reliable transmission strategies with high mobility.
\end{IEEEbiography}

\begin{IEEEbiography}[{\includegraphics[width=1in,height=1.25in,clip,keepaspectratio]{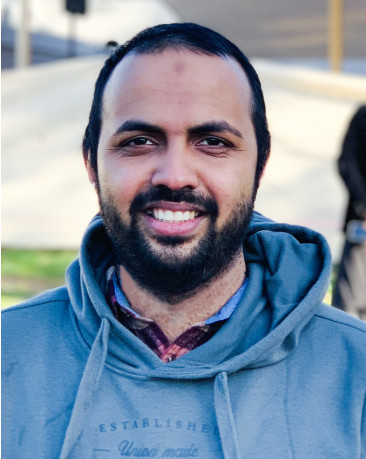}}]{Akram Shafie}
(Member, IEEE) received the Ph.D. degree in Engineering and Computer Science from the Australian National University, Australia, in 2023, and the B.Sc. degree in Electrical and Electronic Engineering from the University of Peradeniya, Sri Lanka, in 2017. He was the recipient of the Best Paper Award at the 2022 IEEE Global Communications Conference (Globecom), and the recipient of the 2019 Sri Lankan President’s Award for Scientific Research. He is currently a Research Associate in the School of Electrical Engineering and Telecommunications at the University of New South Wales, Australia. His research interests include multicarrier modulation for high mobility scenarios and/or underwater communications, terahertz communications, and integrated sensing and communications.
\end{IEEEbiography}

\begin{IEEEbiography}[{\includegraphics[width=1in,height=1.25in,clip,keepaspectratio]{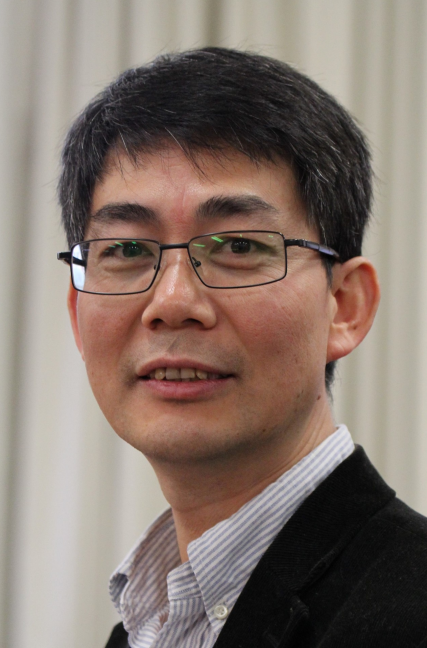}}]{Jinhong Yuan} (Fellow, IEEE) received the B.E. and Ph.D. degrees in electronics engineering from the Beijing Institute of Technology, Beijing, China, in 1991 and 1997, respectively. From 1997 to 1999, he was a Research Fellow with the School of Electrical Engineering, University of Sydney, Sydney, Australia. In 2000, he joined the School of Electrical Engineering and Telecommunications, University of New South Wales, Sydney, Australia, where he is currently the Acting Head of School. He has published two books, five book chapters, over 300 papers in telecommunications journals and conference proceedings, and 50 industrial reports. He is a co-inventor of one patent on MIMO systems and four patents on low-density-parity-check codes. He has co-authored four Best Paper Awards and one Best Poster Award, including the Best Paper Award from IEEE Global Communications Conference, Kuala Lumpur, Malaysia, in 2023, the IEEE International Conference on Communications, Kansas City, USA, in 2018, the Best Paper Award from IEEE Wireless Communications and Networking Conference, Cancun, Mexico, in 2011, and the Best Paper Award from the IEEE International Symposium on Wireless Communications Systems, Trondheim, Norway, in 2007. He is an IEEE Fellow and currently serving as an Associate Editor for the IEEE Transactions on Wireless Communications and IEEE Transactions on Communications. He served as the IEEE NSW Chapter Chair of Joint Communications/Signal Processions/Ocean Engineering Chapter during 2011-2014 and served as an Associate Editor for the IEEE Transactions on Communications during 2012-2017. His current research interests include error control coding and information theory, communication theory, and wireless communications.
\end{IEEEbiography}

\begin{IEEEbiography}[{\includegraphics[width=1in,height=1.25in,clip,keepaspectratio]{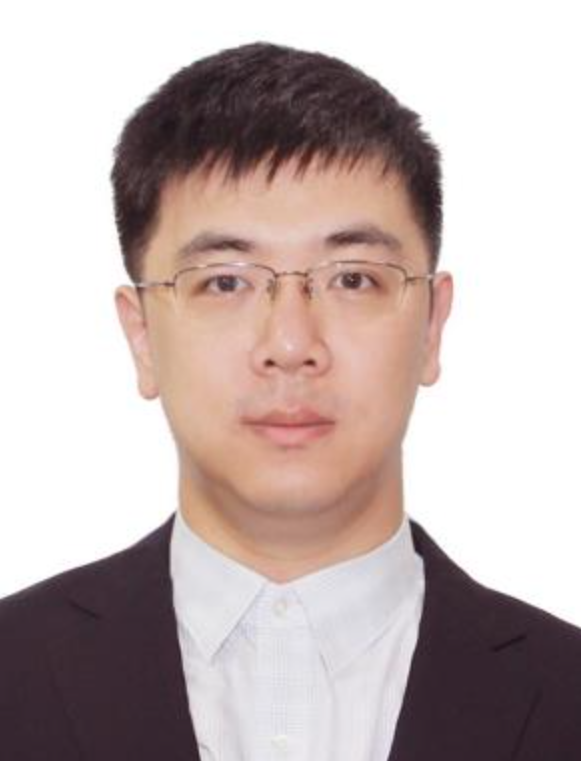}}]{Guoyu Ma}
(Member, IEEE) received the B.S. and Ph.D. degrees in electrical engineering from Beijing Jiaotong University, Beijing, China, in 2012 and 2019 respectively. Currently, he is an associate professor at the School of Electronic and Information Engineering, Beijing Jiaotong University, Beijing, China. His current research interests include the field of machine-type communication and random access.
\end{IEEEbiography}

\begin{IEEEbiography}[{\includegraphics[width=1in,height=1.25in,clip,keepaspectratio]{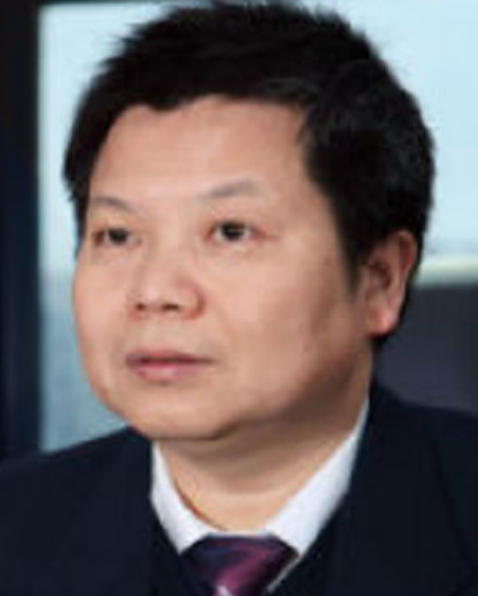}}]{Zhangdui Zhong}(Fellow, IEEE) received the B.E. and M.S. degrees from Beijing Jiaotong University, Beijing, China, in 1983 and 1988, respectively. He is currently a Professor and an Advisor to Ph.D. candidates with Beijing Jiaotong University. He is also the Director of the Innovative Research Team of the Ministry of Education, Beijing. His research interests include wireless communications for railways, control theory and techniques for railways, and GSM-R systems. His research has been widely used in railway engineering, such as Qinghai-Xizang railway, Datong-Qinhuangdao Heavy Haul railway, and many high-speed railway lines in China. He has authored or co-authored seven books, five invention patents, and over 200 scientific research papers in his research area. He is an Executive Council Member of the Radio Association of China, Beijing, and the Deputy Director of the Radio Association, Beijing. He received the Mao Yisheng Scientific Award of China, Zhan Tianyou Railway Honorary Award of China, and the Top 10 Science/Technology Achievements Award of Chinese Universities.
\end{IEEEbiography}

\begin{IEEEbiography}[{\includegraphics[width=1in,height=1.25in,clip,keepaspectratio]{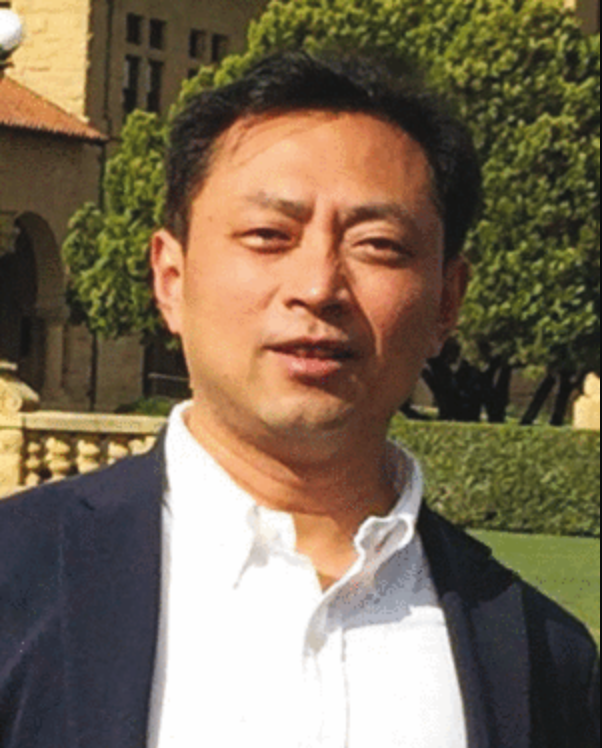}}]{Bo Ai}
 (Fellow, IEEE) received the master's and Ph.D. degrees from Xidian University, China. He received the Honor of Excellent Post-Doctoral Research Fellow from Tsinghua University in 2007. He was a Visiting Professor with the Electrical Engineering Department, Stanford University, Stanford, CA, USA, in 2015. He is currently a Full Professor with Beijing Jiaotong University, where he is the Dean of the School of Electronic and Information Engineering. He is one of the directors for Beijing ``Urban Rail Operation Control System'' International Science and Technology Cooperation Base, and the Backbone Member of the Innovative Engineering based jointly granted by the Chinese Ministry of Education and the State Administration of Foreign Experts Affairs. He is the research team leader of 26 national projects. He holds 26 invention patents. His research interests include the research and applications of channel measurement and channel modeling and dedicated mobile communications for rail traffic systems. He has authored or co-authored eight books and authored over 300 academic research articles in his research area. Five papers have been the ESI highly cited paper. He has won some important scientific research prizes. He has been notified by the Council of Canadian Academies that based on the Scopus database, he has been listed as one of the top 1$\%$ authors in his field all over the world. He has also been feature interviewed by the IET Electronics Letters. Dr. Ai is a fellow of The Institution of Engineering and Technology and an IEEE VTS Distinguished Lecturer. He received the Distinguished Youth Foundation and Excellent Youth Foundation from the National Natural Science Foundation of China, the Qiushi Outstanding Youth Award by the Hong Kong Qiushi Foundation, the New Century Talents by the Chinese Ministry of Education, the Zhan Tianyou Railway Science and Technology Award by the Chinese Ministry of Railways, and the Science and Technology New Star by the Beijing Municipal Science and Technology Commission. He is an IEEE VTS Beijing Chapter Vice Chair and an IEEE BTS Xi' an Chapter Chair. He was a co-chair or a session/track chair of many international conferences. He is an Associate Editor of the IEEE Antennas and Wireless Propagation Letters and the IEEE Transactions on Consumer Electronics, and an Editorial Committee Member of the Wireless Personal Communications Journal. He is the Lead Guest Editor of Special Issues on the IEEE Transactions on Vehicular Technology, the IEEE Antennas and Propagations Letters, and the International Journal on Antennas and Propagations.
\end{IEEEbiography}

\end{document}